\documentclass[11pt]{article}
\pdfoutput=1
\usepackage[centertags]{amsmath}
\usepackage[square, comma, sort&compress,numbers]{natbib}
\usepackage{array,multirow}
\numberwithin{equation}{section}
\usepackage{amssymb}
\usepackage{graphicx}
\usepackage{color}
\usepackage[normalem]{ulem}

\usepackage{epsfig}
\usepackage{bbold}

\usepackage{wrapfig}
\usepackage{float}
\usepackage{soul}
\usepackage{url}

\usepackage{tikz,pgf}
\usetikzlibrary{shapes}
\usetikzlibrary{calc}
\usetikzlibrary{decorations.pathmorphing}
\usetikzlibrary{decorations.pathreplacing,shapes.misc}
\usetikzlibrary{positioning}
\usetikzlibrary{arrows}
\usetikzlibrary{decorations.markings}
\usetikzlibrary{shadings}

\usetikzlibrary{intersections}

\def\sideremark#1{\ifvmode\leavevmode\fi\vadjust{\vbox to0pt{\vss
 \hbox to 0pt{\hskip\hsize\hskip1em
 \vbox{\hsize3cm\tiny\raggedright\pretolerance10000
  \noindent #1\hfill}\hss}\vbox to8pt{\vfil}\vss}}}

\def\be{\begin{equation}}
\def\ee{\end{equation}}
\def\ba{\begin{array}}
\def\ea{\end{array}}

\def\dps{\displaystyle}

\def\tr{{\rm Tr}}

\advance\textheight\topskip
\renewcommand{\tilde}{\widetilde}
\renewcommand{\hat}{\widehat}

\newcommand{\bref}[1]{\textbf{\ref{#1}}}

\renewcommand{\geq}{\,{\geqslant}\,}
\renewcommand{\leq}{\,{\leqslant}\,}

\newcommand{\binner}[2]{%
  {\langle}\kern-4.15pt{\langle}#1{,}\,#2{\rangle}\kern-4.15pt{\rangle}}

\newcommand{\half}{\mathchoice{%
    \ffrac{1}{2}}{\frac{1}{2}}{\frac{1}{2}}{\frac{1}{2}}}

\newcommand{\ffrac}[2]{\raisebox{.5pt}%
  {\footnotesize$\displaystyle\frac{#1}{#2}$}\kern1pt}

\def\cA{\mathcal{A}}

\def\cD{\mathcal{D}}

\def\cF{\mathcal{F}}

\def\cL{\mathcal{L}}
\def\cM{\mathcal{M}}

\def\cO{\mathcal{O}}
\def\cP{\mathcal{P}}

\def\cV{\mathcal{V}}
\def\cW{\mathcal{W}}

\numberwithin{equation}{section} \makeatletter
\@addtoreset{equation}{section}

\def\be{\begin{equation}}
\def\ee{\end{equation}}
\def\ba{\begin{array}}
\def\ea{\end{array}}

\def\dps{\displaystyle}
\def\tr{{\rm Tr}}

\def\tDelta{\tilde\Delta}

\newdimen\tableauside\tableauside=1.0ex
\newdimen\tableaurule\tableaurule=0.4pt
\newdimen\tableaustep
\def\phantomhrule#1{\hbox{\vbox to0pt{\hrule height\tableaurule
width#1\vss}}}
\def\phantomvrule#1{\vbox{\hbox to0pt{\vrule width\tableaurule
height#1\hss}}}
\def\sqr{\vbox{%
  \phantomhrule\tableaustep

\hbox{\phantomvrule\tableaustep\kern\tableaustep\phantomvrule\tableaustep}%
  \hbox{\vbox{\phantomhrule\tableauside}\kern-\tableaurule}}}
\def\squares#1{\hbox{\count0=#1\noindent\loop\sqr
  \advance\count0 by-1 \ifnum\count0>0\repeat}}
\def\tableau#1{\vcenter{\offinterlineskip
  \tableaustep=\tableauside\advance\tableaustep by-\tableaurule
  \kern\normallineskip\hbox
    {\kern\normallineskip\vbox
      {\gettableau#1 0 }%
     \kern\normallineskip\kern\tableaurule}%
  \kern\normallineskip\kern\tableaurule}}
\def\gettableau#1 {\ifnum#1=0\let\next=\null\else
  \squares{#1}\let\next=\gettableau\fi\next}

\def\cA{\mathcal{A}}

\def\cD{\mathcal{D}}

\def\cF{\mathcal{F}}

\def\cL{\mathcal{L}}
\def\cM{\mathcal{M}}

\def\cO{\mathcal{O}}
\def\cP{\mathcal{P}}

\def\cV{\mathcal{V}}
\def\cW{\mathcal{W}}

\numberwithin{equation}{section} \makeatletter
\@addtoreset{equation}{section}

\def\be{\begin{equation}}
\def\ee{\end{equation}}
\def\ba{\begin{array}}
\def\ea{\end{array}}

\def\dps{\displaystyle}

\def\ba{\begin{array}}
\def\ea{\end{array}}

\def\dps{\displaystyle}

\def\sl2{sl(2,\mathbb{R})}
\def\slC2{sl(2,\mathbb{C})}

\def\3j{3$j$}

\def\pref1{\varkappa_{_{j_p,j_1}}}
\def\a{\Omega}
\def\I{I}

\def\centerarc[#1](#2)(#3:#4:#5);%
    {
    \draw[#1]([shift=(#3:#5)]#2) arc (#3:#4:#5);
    }

\usepackage{jheppub}

\makeatletter
\def\@fpheader{\vspace{-.1cm}}
\makeatother

\title{More on Wilson toroidal networks and torus  blocks}

\author[a,b]{Konstantin Alkalaev}

\author[c\,\dagger]{and Vladimir  Belavin}
\note[$\dagger$]{On leave from Lebedev Physical Institute and Institute for Information Transmission Problems, Moscow, Russia.}

\affiliation[a]{I.E. Tamm Department of Theoretical Physics, \\P.N. Lebedev Physical
Institute,\\ Leninsky ave. 53, 119991 Moscow, Russia}
\affiliation[b]{Department of General and Applied Physics, \\
Moscow Institute of Physics and Technology, \\
7 Institutskiy per., Dolgoprudnyi, 141700 Moscow region, Russia}
\affiliation[c]{Physics Department, Ariel University, Ariel 40700, Israel}
\emailAdd{alkalaev@lpi.ru}
\emailAdd{vlbelavin@gmail.com}

\abstract{We consider  the Wilson line networks of the Chern-Simons $3d$ gravity theory with toroidal boundary conditions which calculate global conformal blocks of degenerate quasi-primary operators in torus $2d$ CFT. After general discussion that summarizes  and further extends   results known in the literature we explicitly obtain the one-point torus block and two-point torus blocks through  particular  matrix elements of toroidal Wilson network operators  in irreducible finite-dimensional representations of $\sl2$ algebra.  The resulting expressions are given in two alternative forms using different ways to treat multiple tensor products of $\sl2$ representations: (1) $3mj$ Wigner symbols and intertwiners of higher valence, (2) totally symmetric tensor products of the fundamental $\sl2$ representation.   
}

\begin{document}

\maketitle
\flushbottom

\section{Introduction}
\label{sec:intro}

Conformal blocks are basic ingredients of conformal field theory  correlation functions, they also play crucial role in the conformal bootstrap program~\cite{Belavin:1984vu,Poland:2018epd}. Recently, CFT$_d$ conformal blocks were interpreted in the AdS${}_{d+1}$/CFT${}_{d}$ correspondence as geodesic (Witten) networks stretched in the asymptotically AdS$_{d+1}$ spaces~\cite{Hartman:2013mia,Fitzpatrick:2014vua,Hijano:2015rla,Fitzpatrick:2015zha,Alkalaev:2015wia,Hijano:2015qja,Hijano:2015zsa,Alkalaev:2015lca,Alkalaev:2015fbw,Banerjee:2016qca,Gobeil:2018fzy,Hung:2018mcn,Alekseev:2019gkl}. The alternative description of conformal blocks  in terms of Wilson lines was extensively studied in  \cite{deBoer:2013vca,Ammon:2013hba,deBoer:2014sna,Hegde:2015dqh,Melnikov:2016eun,Bhatta:2016hpz,Besken:2017fsj,Hikida:2017ehf,Hikida:2018eih,Hikida:2018dxe,Besken:2018zro,Bhatta:2018gjb,DHoker:2019clx,Castro:2018srf,Kraus:2018zrn,Hulik:2018dpl,Castro:2020smu,Chen:2020nlj}.\footnote{See also further extensive  developments of the block/network correspondence in different context like black holes \cite{Anous:2016kss,Chen:2017yze,Chen:2018qzm}, heavy-light approximations and other backgrounds \cite{Belavin:2017atm,Kusuki:2018wcv,Kusuki:2018nms,Hijano:2018nhq,Anous:2019yku,Alkalaev:2019zhs,Chen:2019hdv,Alkalaev:2020kxz,Cardona:2020cfy}, supersymmetric extensions \cite{Chen:2016cms,Alkalaev:2018qaz}, higher-point blocks \cite{Hulik:2016ifr,Rosenhaus:2018zqn,Alkalaev:2018nik,Fortin:2019zkm,Parikh:2019ygo,Jepsen:2019svc,Anous:2020vtw}, torus (thermal) CFT \cite{Alkalaev:2016ptm,Kraus:2017ezw,Alkalaev:2017bzx,Gobeil:2018fzy}, etc. }

On the other hand, there is an intriguing relation between the space of quantum states in the three-dimensional Chern-Simons theory in the presence of the Wilson lines and the space of conformal blocks in two-dimensional conformal field theory  noticed a long ago \cite{Witten:1988hf,Verlinde:1989ua,Labastida:1989wt}. Since the $SO(2,2)$ Chern-Simons theory describes $3d$ gravity with the cosmological term then the above relation acquires a new meaning in the context of the AdS$_3$/CFT$_2$ correspondence  \cite{Bhatta:2016hpz,Besken:2016ooo,Fitzpatrick:2016mtp,Kraus:2017ezw,Bhatta:2018gjb}.

The Wilson line networks under consideration   are  typical Penrose's spin networks \cite{Penrose,Baez:1994hx}. Formally, such a network is a graph in AdS space with a number of boundary endpoints, edges associated with $\sl2$ representations and vertices given by 3-valent intertwiners. For a fixed background gravitational connection the Wilson line network is a gauge covariant functional of associated representations. To gain the conformal block interpretation one calculates the matrix element of the network operator between specific boundary states which are  highest(lowest)-weight  vectors in the respective $\sl2$ representations.\footnote{More generally, one can consider arbitrary matrix elements that we call {\it vertex } functions. In Section \bref{sec:further} we show that these are related to correlation functions of descendant operators.     }    

In this paper we revisit the holographic relation between the Wilson line networks and conformal blocks focusing on the case of finite-dimensional $\sl2$ representations. Our primary interest are toroidal Wilson networks in the thermal AdS$_3$ space and corresponding torus blocks. We formulate and calculate one-point and two-point Wilson network functionals which are dual to one-point and two-point torus conformal blocks for degenerate quasi-primary operators. The paper is organised  as follows:

-- in Section \bref{sec:wilson} we review what is known about Wilson networks and how they compute conformal blocks. Here, we briefly recall some necessary  background about Chern-Simons description of $3d$ gravity with the cosmological term.  
Then, on the basis of the findings of Refs. \cite{Bhatta:2016hpz,Besken:2016ooo,Kraus:2017ezw}, we attempt to rethink the whole approach focusing on key elements that would allow one to study higher-point conformal blocks of (quasi-)primary and secondary operators as well as extension to toroidal Wilson networks which are dual to torus conformal blocks.  

-- in Section \bref{sec:toroidal} we define toroidal Wilson network operators with one and two boundary attachments. They are the basis for explicit calculations of one-point blocks and two-point blocks in two OPE channels in the following sections. 

-- in Section  \bref{sec:one-point} we consider torus conformal blocks for degenerate quasi-primary operators which are dual to the Wilson networks carrying finite-dimensional representations of the gauge algebra. 

-- Section \bref{sec:gauge} contains explicit calculation of the one-point toroidal Wilson network operator in two different representations, using \3j Wigner symbols and symmetric tensor product representation. In particular, in the later representation we find  the character decomposition of one-point torus block for degenerate operators. 

-- Section \bref{sec:Two-point} considers explicit calculations of two-point Wilson toroidal networks. In Sections \bref{sec:2s} and \bref{sec:2t} we  formulate the symmetric tensor product representation of the toroidal Wilson networks. Explicit demonstration that the corresponding network operators calculate 2-point blocks is given by one simple example (unit conformal weights)  for each OPE channel contained in Appendix \bref{app:ex}. In Sections \bref{S2pt-proof} and \bref{T2pt-proof} we explicitly calculate the $s$-channel and $t$-channel toroidal networks for general conformal weights using \3j Wigner symbols and show that the resulting functions coincide with 2-point $s$-channel and $t$-channel torus blocks.

-- concluding remarks and future perspectives  are shortly discussed in Section \bref{sec:concl}. Technical details are collected in Appendices \bref{sec:sl2}--\bref{app:ex}.

\section{Wilson networks vs conformal blocks} 
\label{sec:wilson}

In this section we mainly  review Wilson line approach to conformal blocks  proposed and studied in different contexts in \cite{Bhatta:2016hpz,Besken:2016ooo,Fitzpatrick:2016mtp,Kraus:2017ezw,Bhatta:2018gjb}. Here, we  rephrase the whole construction in very general terms  underlying key elements that finally allow direct passing from concrete calculations of sphere blocks in the above references to calculation of torus blocks. We will discuss only the case of (non-unitary) finite-dimensional $\sl2$ representations (see Appendix \bref{sec:sl2}). The Wilson networks carrying (unitary) infinite-dimensional representations and the corresponding global sphere blocks are considered in  \cite{Bhatta:2016hpz,Fitzpatrick:2016mtp}.

\subsection{Brief review of $3d$ Chern-Simons gravity theory}
\label{sec:CS}

The Chern-Simons formulation of $3d$  gravity with the cosmological term is obtained by combining the dreibein and spin-connection into the $o(2,2)$-connection $\cA$ \cite{Achucarro:1987vz,Witten:1988hc} (for extensive review see e.g. \cite{Banados:1998gg,Ammon:2013hba}). Decomposing the gauge algebra as $o(2,2) \approx \sl2\oplus \sl2$ one introduces associated (anti)-chiral connections $A, \bar A$ in each simple factor $\sl2$ with basis elements $J_{0,\pm1}$, 
see the Appendix \bref{sec:sl2} for more details. Then, the 3d gravity action is given by  the $o(2,2)$ Chern-Simons action
\be
\label{action}
S[\cA]=\frac{k}{4\pi} \int_{\cM^3} \tr\big(\cA\wedge \cA +\frac{2}{3} \cA\wedge \cA\wedge \cA\big) \;,
\ee
where $k$ is related to the $3$-dimensional Newton constant $G_3$ through $k = l/(4G_3)$ and $l$ is the AdS$_3$ radius, and $\tr$   stands for the Killing invariant form. Equivalently, the action can be factorized as $S[\cA] = S[A]-S[\bar{A}]$, where each chiral component is the $\sl2$ Chern-Simons action. The convenient choice of local  coordinates is given by  $x^\mu=(\rho,z, \bar z)$ with radial  $\rho\geq 0$ and (anti)holomorphic $z, \bar z \in \mathbb{C}$. 
 
The equations of motion that follow from the CS action \eqref{action} are generally solved by flat $o(2,2)$-connections $\cA$. After imposing appropriate boundary conditions the solutions yielding   flat boundary metric can be written as the gauge transformed (chiral) connection $A = U^{-1} \a U+ U^{-1} dU$ with  the gauge group element $U(x) = \exp{\rho J_0}$ \cite{Banados:1994tn} and the holomorphic gravitational connection given by 
\be
\label{con_a}
\a=\bigg(J_1-2\pi\frac{6T(z)}{c} J_{-1}\bigg) dz\;,
\ee 
where $T(z)$ is the holomorphic boundary stress tensor, the central charge $c$ is defined through the Brown-Henneaux relation $c = 3l/(2G_3)$ \cite{Brown:1986nw}. The same anti-holomorphic connection $\bar \Omega = \bar \Omega({\bar z})$ arises in the anti-chiral  $\sl2$ sector.

Considering a path $L$ connecting two points $x_1, x_2\in\cM_3$ we can associate to $L$ the following chiral Wilson line operators
\be
\label{wilson}
W_R[L] =  \cP \exp{\left(-\int_L \Omega\right)}\;,
\ee
where the chiral $\sl2$ connection is given by \eqref{con_a} in some representation $R$. Similarly, one can consider $\overline{W}_R[L]$ in the anti-chiral sector.   Under the gauge group,  the Wilson operator transforms homogeneously as $W_R[L] \to U_R(x_2) W_R[L] U_{R}^{-1}(x_1)$, where the gauge group elements are $U_R = \exp \epsilon  J_R$ with generators $J_R$ in the representation $R$. As we deal with the flat connections, the Wilson line operators depend only on the path $L$ endpoints and  on the topology of the base manifold $\cM_3$.  (Anti-)chiral Wilson operators \eqref{wilson} are instrumental when discussing  (anti-)holomorphic conformal blocks in the boundary conformal theory.

\subsection{General construction}
\label{sec:general}

The  Euclidean AdS$_3$ space metric can be obtained from \eqref{con_a} by taking a  constant boundary stress tensor. In what follows we  discuss spaces with both periodic and non-periodic time directions. In the non-periodic case, the stress tensor can be chosen as $T(z) = 0$ so that  the  chiral gravitational connection \eqref{con_a} takes the form 
\be
\label{con-glob-12}
\a= J_1 dz\;,
\ee
and the corresponding  AdS$_3$ metric is given in the Poincare coordinates. In  the periodic case (thermal AdS$_3$), the stress tensor  is $T(z) = -c/48\pi$  so that the chiral connection is given by 
\be
\label{con-glob-13}
\a=\bigg(J_1+\frac{1}{4} J_{-1}\bigg) dw\;,
\ee
along with the standard identifications $w \sim w+2\pi$ and $w \sim w + 2\pi \tau$, where $i\tau\in \mathbb{R}_-$. The boundary (rectangular) torus is defined by  the modular parameter $\tau$  while the conformal symmetry algebra in the large-$c$ limit is contracted to the finite-dimensional $\sl2\oplus\sl2$. 

In the chiral sector, the Wilson line \eqref{wilson} for the connections \eqref{con-glob-12} or \eqref{con-glob-13} is the holonomy of the chiral gauge field along the path $L$ with endpoints $x_1$ and $x_2$: 
\be
\label{chiral_wilson}
W_{a}[x_1,x_2]=\cP\exp \bigg(-\int_{x_1}^{x_2} \a \bigg) = \exp \left(x_{12}\, \a\right)\;,
\ee
where $x_{mn} = x_m-x_n$, and $a$ labels a finite-dimensional spin-$j_a$ representation $\cD_a$ of the chiral gauge algebra $\sl2$. Recall that the Wilson line operators have the transition property  
\be
\label{trans}
W_{a}[x_1,x_2] = W_{a}[x_1,x] W_{a}[x,x_2]\;,
\ee
where $x$ is some intermediate point. The relation \eqref{trans} is obvious for coordinate independent connections like \eqref{con-glob-12} and \eqref{con-glob-13}. 
\vspace{2mm}

\noindent In order to realize conformal blocks through the Wilson networks we need the following ingredients. 
\begin{itemize}

\item[1)] The Wilson line $W_{a}[z,x]$ in a spin-$j_a$ representation $\cD_a$ of $\sl2$ algebra,    connecting the external operator $\cO_{\Delta_a}(z, \bar z)$  on the boundary with some point $x$ in the bulk. The conformal dimension of the boundary operator is $\Delta_a = -j_a$.

\item[2)]The Wilson line  $W_{a}[x,y]$ connecting two bulk points $x$ and $y$. In the thermal AdS$_3$, the thermal cycle yields the Wilson loop $W_{\alpha}[x,x+2\pi \tau]$.

\item[3)]  The trivalent vertex in the bulk point $x_b$ connects three Wilson line operators associated with three representations $\cD_{b}, \cD_{c}$, and $\cD_{a}$ by means of the 3-valent  intertwiner operator 
\be
\label{inter1}
\I_{a; b, c}:\qquad  \cD_b \otimes \cD_c \to \cD_a\;,
\ee
which satisfies the defining $\sl2$ invariance property 
\be
\label{inter2}
\I_{a; b,c}\, U_b\,  U_c = U_a\, \I_{a; b,c}\;,
\ee
where $U_{\alpha}$ labelled by $\alpha=a, b,c$ are  linear operators acting in the respective representation spaces. In other words, the intertwiner spans the one-dimensional space of $\sl2$ invariants Inv$(\cD^*_a \otimes \cD_b \otimes \cD_c)$, where $*$ denotes  a contragredient representation.  

\item[4)] The Wilson line attached to the boundary acts on a particular state $|a\rangle \in \cD_a$. 
\end{itemize}

In general, $n$-point global conformal blocks $\cF(\Delta, \tDelta|{\bf q},{\bf z})$ on Riemann surface of genus $g$ with  modular parameters ${\bf q} = (q_1,..., q_{g})$, with external and intermediate conformal dimensions $\Delta$ and $\tilde \Delta$, can be calculated as the following matrix element 
\be
\label{Phi}
\cF(\Delta_i, \tDelta_j|{\bf q},{\bf z}) = \langle\!\langle\,\Phi\left[W_{a}, \I_{b;c, d}|{\bf q},{\bf z}\right]\,\rangle\!\rangle\;.
\ee 
Here, the {\it Wilson network operator} $\Phi[W_{a}, \I_{b;c, d}]$ is built of Wilson line   operators $W_{a}$ associated to a particular (bulk-to-bulk or bulk-to-boundary) segments joined together by 3-valent intertwiners $\I_{b;c, d}$ to form a network with the boundary endpoints ${\bf z}=(z_1,...,z_n)$. The double brackets mean that one calculates a particular matrix element of the operator $\Phi$ between specific vectors of associated  $\sl2$ representations in such a way that the resulting quantity is $\sl2$ gauge algebra  singlet. Using general arguments one may show that the matrix element \eqref{Phi}: (a)  does not depend on positions of bulk vertex points due to the gauge covariance of the Wilson operators, (b) transforms under $sl(2)$ conformal boundary transformations as $n$-point correlation function.

\subsection{Vertex functions}

In what follows we discuss  examples of the operator \eqref{Phi}: 2-point, 3-point and 4-point Wilson networks in the AdS$_3$ space with the spherical (plane) boundary.  Let us consider first the trivalent vertex consisting of three boundary anchored Wilson lines meeting in the bulk point $x$. Let $|a\rangle$ be some vector in the spin-$j_a$ representation $\cD_a$ that we call a boundary vector. Acting with the bulk-to-boundary Wilson line $W_a[x,z]$ we can obtain the following bra and ket vectors 
\be
\label{a_tilde}
\ba{c}
|\tilde a\rangle = W_a[x,z]|a\rangle\;,
\\
\\
\langle\tilde a| = \langle a| W_a[z,x]\;,
\ea
\ee 
to be associated with some quasi-primary or secondary boundary operator $\cO(z,\bar z)$ belonging to the conformal family $[\cO_{\Delta_a}]$ of dimension $\Delta_a = -j_a$. 

Bra and ket vectors \eqref{a_tilde} are the only elements of the Wilson network operator \eqref{Phi} which depend on positions of boundary operators. Thus, it is their properties that completely define how the resulting CFT correlation function (block) transforms with respect to the global conformal symmetry algebra. One can show that depending on the choice of particular $W_a$ and $|a\rangle \in \cD_a$, the conformal invariance of the correlation function of quasi-primary operators is guaranteed by the following basic property \cite{Besken:2016ooo,Kraus:2017ezw,Fitzpatrick:2016mtp} 
\be
\label{conf_trans}
\left(\cL_n-C_n{}^m J_m\right)|\tilde a\rangle = 0\;,
\qquad n=0,\pm1\;,
\qquad 
\ee  
where $[C_n{}^m]$ is some $[3\times3]$ constant matrix.  
It claims that holomorphic conformal transformation is generated by a combination of the chiral $\sl2$ gauge transformations.  Here, $J_n$ are (chiral) $\sl2$ gauge algebra  generators taken in the representation $\cD_a$ and $\cL_n$ are the boundary conformal generators represented by differential operators in coordinates $z$, satisfying (holomorphic) $\sl2$ conformal algebra commutation relations $[\cL_m,\cL_n]=(m-n)\cL_{m+n}$. Explicit form of $C_n{}^m$ is fixed by particular choice of the gravitational connections defining $W_a$ and boundary vectors $|a\rangle$ (see below).\footnote{Technically, in this paper, the matrix $C$ is calculated case by case and, moreover, just for two background gravitational connections \eqref{con-glob-12} and \eqref{con-glob-12}, both with constant coefficients. It would be important to formalize its possible properties like unitarity, etc. (The only obvious property now is that $C$ is invertible.) On the other hand, conceptually, it is obvious that the matrix $C$ is a derived object. Its exact definition and properties can be rigorously obtained from the holographic Ward identities of dual $3d$ Chern-Simons theory and CFT$_2$ along the lines discussed in the Appendix A of Ref. \cite{Fitzpatrick:2016mtp}. }        

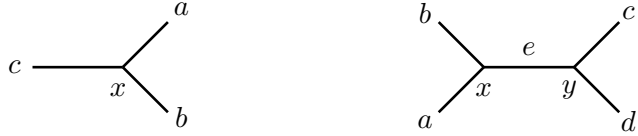
\begin{figure}[H]
\centering

\begin{tikzpicture}[line width=1pt,scale=0.60]

\draw (-20,-1) -- (-18,-1);

\draw (-18,-1) -- (-17,-2);
\draw (-18,-1) -- (-17,0);

\draw (-20.4,-1.0) node {$c$};
\draw (-16.7,.3) node {$a$};
\draw (-16.7,-2.1) node {$b$};

\draw (-18.1,-1.5) node {$x$};


\draw (-10,-1) -- (-8,-1);
\draw (-10,-1) -- (-11,-2);
\draw (-10,-1) -- (-11,0);
\draw (-8,-1) -- (-7,-2);
\draw (-8,-1) -- (-7,0);

\draw (-11.3,0.2) node {$b$};
\draw (-11.3,-2.2) node {$a$};
\draw (-6.8,-2.2) node {$d$};
\draw (-6.8,0.2) node {$c$};
\draw (-9.0,-0.6) node {$e$};
\draw (-10.0,-1.5) node {$x$};
\draw (-8.1,-1.5) node {$y$};


\end{tikzpicture}

\caption{Wilson networks: trivalent vertex (left) and four-valent vertex given by two trivalent vertices joined by an edge (right). } 
\label{34net}
\end{figure}

\paragraph{3-point vertex function.} Following the general definition of the Wilson network operator \eqref{Phi} we use the intertwiner \eqref{inter1} and introduce a trivalent {\it vertex} function  (see Fig. \bref{34net}) as the following matrix element     
\be
\label{tri1}
V_{a, b, c}({\bf z}) = \langle\tilde a| \,\I_{a; b, c}\, |\tilde b\rangle \otimes |\tilde c\rangle = \langle a| W_a[z_1,x] \,\I_{a; b, c}\, W_b[x,z_2] W_c[x,z_3] | b\rangle \otimes | c\rangle\;,
\ee 
where ${\bf z} = (z_1, z_2, z_3)$ stands for positions of boundary conformal operators, $| a\rangle, | b\rangle, | c\rangle$ are arbitrary boundary vectors. Using the invariance condition  \eqref{inter2} in the form    
\be
\label{tri20}
\I_{a; b, c}\, W_b[x,z_2] = W_a[x,z_2]\I_{a; b, c}\, W^{-1}_c[x,z_2] = W_a[x,z_2]\I_{a; b, c}\, W_c[z_2,x]\;,
\ee
along with the transition property \eqref{trans} the trivalent vertex function can be represented as 
\be
\label{tri2}
V_{a, b, c}({\bf z}) =  \langle a| \, W_a[z_1,z_2] \,\I_{a; b, c}\, W_c[z_2,z_3] \,| b\rangle \otimes | c\rangle\;.
\ee
This expression  can be equivalently obtained by choosing the bulk vertex point $x = z_2$ yielding $W_b[z_2,z_2]=\mathbb{1}$. This is legitimate since we noted earlier that the resulting Wilson network does not depend on location of bulk vertices.  On the other hand, this freedom in choosing the vertex point is encoded in the intertwiner transformation property.

Two comments are in order. First, in order to have a non-trivial intertwiner with the property \eqref{inter2} the weights of three representations must satisfy the triangle inequality. Indeed, tensoring two irreps as in \eqref{inter1} we find out the Clebsch-Gordon  series  
\be
\label{fusion}
\cD_{a} \otimes \cD_{b}= \bigoplus_{j_c=|j_a-j_b|}^{j_a+j_b} \cD_{c}\;.
\ee 
If a representation $\cD_c$ of a given spin $j_c$  arises in the Clebsch-Gordon series then the intertwiner is just a projector, otherwise it is zero. Equivalently, $j_a+j_b- j_c \geq 0$. Second, one may rewrite \eqref{tri2} through the matrix elements in the standard basis \eqref{standard} by inserting  resolutions of identities,   
\be
\label{resolve}
\mathbb{1} = \sum_{m=-j}^j |j,m\rangle \langle j, m|
\ee
to obtain 
\be
\label{tri3}
V_{a, b, c}({\bf z}) =  \sum_m \sum_k\sum_n\, \Big( \langle j_a,m|\,\I_{a; b, c}\,  | j_b,k\rangle \otimes |j_c,n\rangle \Big)\langle \tilde a|j_a, m\rangle \langle j_b,k| \tilde b\rangle \langle j_c, n|  \tilde c\rangle \;.
\ee  
In this form the trivalent vertex  function is represented  as  a product of four matrix elements which can be drastically simplified when $|a\rangle, | b\rangle$, and $|c\rangle$ are chosen in some canonical way like lowest-weight or highest-weight  vectors. The last three factors are matrix elements of the Wilson operators, or, equivalently, coordinates of tilded vectors in the standard basis.  The first factor is the matrix element of the intertwiner which in fact is the Wigner \3j symbol.\footnote{Strictly speaking, we consider here $SL(2, \mathbb{R})\approx SU(1,1)$ Wigner $3mj$ symbols which are generally different from $SU(2)$ Wigner $3mj$ symbols for arbitrary (unitary or non-unitary, finite- or infinite-dimensional) representations. However, in this paper we deal only with finite-dimensional representations for which these two types of symbols are identical \cite{HOLMAN19661}. Note also that if we consider Wilson networks in Euclidean dS$_3$  gravity \cite{Castro:2020smu} where the spacetime isometry group is $SO(4) \approx SU(2)\times SU(2)$, then  we can directly apply the standard Wigner \3j symbol calculus. } Indeed, let us denote the matrix element of the intertwiner and \3j symbol as 
\be
\label{mat_int}
[I_{a;b,c}]^m{}_{kn} = \langle j_a,m|\,\I_{a; b, c}\,  | j_b,k\rangle \otimes |j_c,n\rangle\;,
\;\;\qquad
[W_{a,b,c}]_{mkn} = 
\begin{pmatrix}
j_a & j_b & j_c \\
m & k & n 
\end{pmatrix}\;.
\ee
Here, each of magnetic numbers $m,n,k$ runs its domain. Then, the two tensors are related as
\be
\label{tri3-1}
[I_{a;b,c}]^m{}_{kn}  = \sum_l\epsilon^{(a)}{}^{ml}  [W_{a,b,c}]_{lkn}\;,
\ee
where $\epsilon^{(a)}{}^{ml}$ is the Levi-Civita tensor in the $\cD_a$ representation. Obviously, both tensors are $\sl2$ invariant, while the \3j symbol spans Inv$(\cD_a \otimes \cD_b \otimes \cD_c)$. The Levi-Civita tensor in $\cD_a$ is given by 
\be
\label{levi_civita}
\epsilon^{(a)}{}^{mn} = (-)^{j_a-m}\delta_{m,-n}  = 
\sqrt{2j_a+1}\begin{pmatrix}
j_a & j_a & 0 \\
m & m & 0 
\end{pmatrix}=
\begin{pmatrix}
j_a \\
mn 
\end{pmatrix}\;.
\ee
The last equality introduces the 1$jm$ Wigner symbol which is considered as an invariant metric relating  the standard and contragredient standard bases. In particular, this object allows introducing 2-point vertex function as
\be
\label{tri1}
V_{a, b}({\bf z}) = \langle\tilde a| \,\I_{a; b}\, |\tilde b\rangle 
=
\delta_{ab} \langle a| \I_{a; a}\, W_a[z_1,z_2] | a\rangle \;,
\ee 
where $\I_{a;a}$ is 2-valent intertwiner belonging to Inv$(\cD_a^* \otimes \cD_a)$ which definition directly follows from \eqref{inter1}, \eqref{inter2} at $\cD_c = \mathbb{1}$. Thus, 
\be
\label{valent2}
[\I_{a; a}]_{m}{}^n = \begin{pmatrix}
j_a \\
mn 
\end{pmatrix}\;.
\ee 

Coming back to the 3-point vertex functions one may explicitly check that choosing the boundary vectors as highest-weight elements of the respective spin-$j_\eta$ representations $\eta=a,b,c$ (see Appendix \bref{sec:sl2})
\be
\label{HW_i}
|\eta\rangle =|\mathbb{hw}\rangle_\eta\;:
\qquad 
J_{-1}|\mathbb{hw}\rangle_\eta = 0\;,
\qquad
J_{0}|\mathbb{hw}\rangle_\eta = j_\eta |\mathbb{hw}\rangle_\eta\;, 
\ee 
along with the Wilson line operator in Euclidean AdS$_3$ space defined by the  connection \eqref{con-glob-12},
one reproduces the 3-point function of quasi-primary operators  on the plane \cite{Bhatta:2016hpz,Besken:2016ooo}:
\be
\label{rel_3}
V_{a, b, c}({\bf z}) = \langle \cO_{\Delta_a}(z_1, \bar z_1)\cO_{\Delta_b}(z_2, \bar z_2)\cO_{\Delta_c}(z_3, \bar z_3) \rangle\;. 
\ee
One can show that the basic transformation property \eqref{conf_trans} guaranteeing the conformal invariance of \eqref{rel_3} is defined by the backward identity matrix $C_{m}{}^{n}$, i.e.,
\be
\label{backward}
\begin{aligned}
&\big(J_1+\cL_{-1}\big)W|\mathbb{hw}\rangle=0\;,\\[1pt]
&\;\;\big(J_0+\cL_{0}\big)W|\mathbb{hw}\rangle=0\;,\\[1pt]
&\big(J_{-1}+\cL_{1}\big)W|\mathbb{hw}\rangle=0\;.
\end{aligned}
\ee

\paragraph{4-point vertex function.} Further, we may consider 4-point vertex function between four representations  $\cD_a, \cD_b, \cD_c, \cD_d$ built as two trivalent vertices attached to each other through an intermediate bulk-to-bulk Wilson line $W_e \equiv W_e[y,x]$ carrying the representation $\cD_e$ (see Fig. \bref{34net}), namely, 
\be
\label{four1}
V_{a,b,c,d|e}({\bf z}) = \langle\tilde d| \,\I_{d; c, e} \, W_e\,\I_{e; a, b}\, |\tilde a\rangle \otimes |\tilde b\rangle \otimes |\tilde c\rangle \;.
\ee
Using the transition property we can represent $W_e[y,x] = W_e[y,0]W_e[0,x]$ and then: (1) for the left factor we repeat arguments around \eqref{tri20} to neglect dependence on $y$, (2) for the right factor we use the intertwiner transformation property to neglect dependence on $x$. The result is that positions $x,y$ fall out of \eqref{four1}. Effectively, it means that we set $x=y=0$ so that the intermediate Wilson line operator trivializes $W_e[x,x] = \mathbb{1}$. All in all, we find that the vertex function can be cast into the form  
\be
\label{four2}
V_{a,b,c,d|e}({\bf z}) = \langle d| W_d[z_4,0] \,\I_{d; c, e}\, \I_{e; a, b}\,    W_a[0,z_1]\,W_b[0,z_2]\,W_c[0,z_3]\,|a\rangle \otimes |b\rangle \otimes |c\rangle\;.
\ee
Similarly to the previous consideration of the trivalent function one may reshuffle  the Wilson operators using the intertwiner transformation property and by inserting the resolutions of identities  represent the final expression as contractions of six matrix elements.  Choosing $|a\rangle, | b\rangle, | c\rangle$, and $|d\rangle$  to be highest-weight vectors in their representations  one  directly finds  4-point conformal block on the sphere \cite{Bhatta:2016hpz,Besken:2016ooo}. 

For our further purposes, the 3-point function \eqref{tri1} or \eqref{tri2} along with the 4-point function \eqref{four1} will prove convenient to build conformal blocks on the torus 
(see Section~\bref{sec:toroidal}). Building the operator $\Phi$ \eqref{Phi} for  the Wilson networks with more endpoints and edges is expected to give higher point conformal blocks on the sphere though  this has not been checked explicitly (except for 5-point sphere block in the comb channel \cite{Bhatta:2016hpz}).  In the next section we discuss $\Phi$ in terms of $n$-valent intertwiners. 

\subsection{Further developments}
\label{sec:further}

Here we extend the general discussion in the previous section by considering some novel features of the Wilson network vertex functions.   

\paragraph{Descendants.} Let us demonstrate that choosing the boundary vectors as descendants of highest-weight vectors we reproduce 3-point function of any three (holomorphic) secondary  operators
\be
\cO_{\Delta}^{(l)}(z,\bar z) = (\cL_{-1})^l \cO_\Delta(z,\bar z)\;,  
\ee     
where $\cL_{-1}$ is one of three conformal generators on the plane, $\cL_n = z^{n+1}\partial+(n+1)\Delta z^n$,  $n = 0, \pm 1$. Taking descendants as
\be
\label{descendants}
|\eta\rangle =|j_\eta, k\rangle = (J_{1})^k|\mathbb{hw}\rangle_\eta\;,
\qquad \eta = a,b,c\;,
\ee
and using that: (1) the gravitational connection is given by \eqref{con-glob-12} so that $[W_a[x,y], J_1] = 0$, (2) the property $J_1 \sim \cL_{-1}$ \eqref{backward},  we find that the respective (holomorphic) 3-point correlation function is given by 
\be
\ba{l}
V_{a, b, c}({\bf z}) = \langle \cO^{(k)}_{\Delta_a}(z_1, \bar z_1)\cO^{(l)}_{\Delta_b}(z_2, \bar z_2)\cO^{(m)}_{\Delta_c}(z_3, \bar z_3) \rangle
\\
\\
\hspace{30mm}=\left(\cL^{(1)}_{-1}\right)^k \left(\cL^{(2)}_{-1}\right)^l \left(\cL_{-1}^{(3)}\right)^m \langle \cO_{\Delta_a}(z_1, \bar z_1)\cO_{\Delta_b}(z_2, \bar z_2)\cO_{\Delta_c}(z_3, \bar z_3) \rangle\;, 
\ea
\ee
where superscript $(i)$ in the last line refers to $z_i$ coordinates.

Similarly,  4-point functions of secondary conformal operators can be obtained by choosing the boundary states to be descendants vectors in the respective representations. Indeed, 4-point correlation function of quasi-primary conformal operators decomposes as
\be
\langle \cO_{\Delta_a}(z_1, \bar z_1)\cO_{\Delta_b}(z_2, \bar z_2)\cO_{\Delta_c}(z_3, \bar z_3)\cO_{\Delta_d}(z_4, \bar z_4) \rangle = \sum_{e, \tilde e} C_{ab,e\tilde e}\, C_{e\tilde e, cd} \,V_{a,b,c,d|e}({\bf z})\,\bar V_{a,b,c,d|\tilde e}\,({\bf \bar z})\;,
\ee  
where $C_{ab,e\tilde e}$ and $C_{e\tilde e, cd}$ are structure constants, $e, \tilde e$ stand for intermediate representations $\cD_e$ and $\cD_{\tilde e}$ in (anti)holomorphic sectors, and $|a\rangle,|b\rangle,|c\rangle,|d\rangle$ inside the vertex functions are boundary highest-weight vectors \eqref{HW_i} as discussed below the 4-point vertex function \eqref{four2}. Then, applying all forgoing arguments we obtain the  4-point correlation function of secondary operators. 

\paragraph{Higher-valent intertwiners.} The 4-point vertex function \eqref{four2} is basically defined by contraction of two intertwiners by one index. The resulting $sl(2)$ invariant tensor is a 4-valent intertwiner,
\be
I^{(1)}_{ab,cd|e} = \I_{d; c, e}\, \I_{e; a, b}\;.
\ee     
Similar to \eqref{mat_int}, using the definition of the Levi-Civita tensor \eqref{levi_civita} we can calculate the 4-valent intertwiner in the standard  basis as 
\be
\left[I^{(1)}_{ab,cd|e}\right]_{n_1n_2n_3}{}^{n_4} = (-)^{j_d-n_4}\sum_m (-)^{j_e-m}
\begin{pmatrix}
j_a & j_b & j_e \\
n_1 & n_2 & m 
\end{pmatrix}
\begin{pmatrix}
j_e & j_c & j_d \\
-m & n_3 & -n_4 
\end{pmatrix}\;.
\ee
 
Fixing the order of $a,b,c,d$ one can introduce one more 4-valent intertwiner with shuffled edges   
\be
I^{(2)}_{ac,bd|e} = \I_{d; b, e}\, \I_{e; a, c}\;,
\ee  
or, in components, 
\be
\left[I^{(2)}_{ac,bd|e}\right]_{n_1n_2n_3}{}^{n_4} = (-)^{j_d-n_4}\sum_m (-)^{j_e-m}
\begin{pmatrix}
j_a & j_c & j_e \\
n_1 & n_3 & m 
\end{pmatrix}
\begin{pmatrix}
j_e & j_b & j_d \\
-m & n_2 & -n_4 
\end{pmatrix}\;.
\ee

The two intertwiners provide two bases in  Inv$(\cD^*_d\otimes \cD_a\otimes \cD_b \otimes \cD_c)$.  In the standard  basis,  one intertwiner is expressed in terms of the other by the following relation 
\be
\label{crossing}
I^{(2)}_{e|ac,bd}  =  \sum_{j_k}  (-)^{j_b+j_c+j_e+j_k}\left(2j_k+1\right)\, 
\begin{Bmatrix}
j_a & j_b & j_k \\
j_d & j_c & j_e
\end{Bmatrix} \,
I^{(1)}_{k|ab,cd}\;,  
\ee   
where, by definition,  the expansion coefficients are given by the 6$j$ Wigner symbol. In terms of the conformal block decomposition of the 4-point correlation function $\langle \cO_{\Delta_a}\cO_{\Delta_b}\cO_{\Delta_c}\cO_{\Delta_d} \rangle$ we say about exchanges in two OPE channels, while the change of basis \eqref{crossing} is the crossing relation. We see that the crossing matrix is given by the 6$j$ Wigner symbol which in its turn can be expressed from \eqref{crossing} as a contraction of two distinct 4-valent intertwiners or, equivalently, four 3-valent intertwiners.\footnote{The Wigner 6$j$ symbols for the conformal group $o(d-1,2)$ have attracted some interest recently for their role in the crossing (kernel) equations and for CFT$_d$ 4-point functions, see e.g. \cite{Ponsot:1999uf,Gadde:2017sjg,Liu:2018jhs,Meltzer:2019nbs,Sleight:2018epi,Albayrak:2020rxh}.}
    
Intertwiners of arbitrary higher valence can be introduced in the same manner to build $n$-point conformal blocks. Fixing the order of representations, an $n$-valent intertwiner can be defined by contracting $n-2$ copies of the 3-valent intertwiner by means of $n-3$ intermediate representations  with representations ordered in different ways,  
\be
\label{n_valent}
I_{a_1,a_2,...,a_n |e_1,...,e_{n-3}} = \I_{a_1; a_2, e_1}\, \I_{e_1; a_3, a_4} ... \I_{e_{n-3};a_{n-1},a_n}\;.
\ee 
Each of possible contractions defines a basis in Inv$(\cD^*_{j_1}\otimes \cdots \otimes \cD_{j_n})$ which can be changed by an appropriate Wigner $3(n-2)j$ symbol. E.g. in the five-point case the crossing matrices are given by Wigner 9$j$ symbol, etc. 

The corresponding $n$-point blocks of conformal (quasi-primary/secondary) operators in $[\cO_{\Delta_i}]$ with dimensions $\Delta_i = -j_{a_i}$ are built by acting with a given $n$-valent intertwiner on $n$ boundary states $|\tilde a_i\rangle=W_{i}[0,z_i]|a_i\rangle$, $i=1,...,n$,  see \eqref{a_tilde},  as
\be
\label{Phi_s}
F(\Delta_i, \tDelta_j|{\bf z}) = \langle \tilde a_1| 
I_{a_1,a_2,...,a_n |e_1,...,e_{n-3}}|\tilde a_2 \rangle \otimes \cdots  \otimes |\tilde a_n\rangle\;,
\ee 
where the intertwiner is built by a particular ordering the representations that corresponds to  given  exchange channels in CFT$_2$ with dimensions $\tDelta_l = -j_{e_l}$, $l=1,...,n-3$.   In this way, we explicitly  obtain the Wilson network operator on the sphere \eqref{Phi}.  

\section{Toroidal Wilson networks}
\label{sec:toroidal}

As discussed  in Section \bref{sec:general},  due to the gauge covariance, the Wilson networks do not depend on positions of vertex points in the bulk. It follows that bulk-to-bulk Wilson lines are effectively shrunk to points so that all diagrams with exchange channels expanded in trees and loops are given by contact diagrams only. However, on non-trivial topologies like (rigid) torus we discuss in this paper, there are non-contractible cycles. Then the associated Wilson networks will contain non-contractible loops given by non-trivial holonomies. 

\begin{figure}[H]
\centering

\begin{tikzpicture}[line width=1pt,scale=0.50]


\draw (-20,0) -- (-18,0);

\draw (-18,0) -- (-17,-1);
\draw (-18,0) -- (-17,1);

\draw (-18,-.5) node {$x$};

\draw (-20.5,0) node {$c$};
\draw (-17,1.5) node {$a$};
\draw (-17,-1.5) node {$b$};

\draw[blue, dashed] (-17,1) .. controls (-14,3) and (-14,-3) .. (-17,-1);


\draw (-9,-1) -- (-7,-1);
\draw (-9,-1) -- (-10,-2);
\draw (-9,-1) -- (-10,0);
\draw (-7,-1) -- (-6,-2);
\draw (-7,-1) -- (-6,0);

\draw (-9,-1.5) node {$x$};

\draw (-7,-1.5) node {$y$};

\draw (-8,-.5) node {$e$};

\draw (-10.5,-2.1) node {$a$};
\draw (-5.6,-2.1) node {$c$};

\draw (-10.5,.2) node {$b$};
\draw (-5.6,.2) node {$d$};

\draw[blue, dashed] (-10,0) .. controls (-12,3) and (-4,3) .. (-6,0);


\draw (0,0) -- (2,0);
\draw (0,0) -- (-1,-1);
\draw (0,0) -- (-1,1);
\draw (2,0) -- (3,-1);
\draw (2,0) -- (3,1);

\draw (0,-.5) node {$x$};

\draw (2,-.5) node {$y$};

\draw (-1.3,1.2) node {$b$};

\draw (3.0,1.4) node {$d$};

\draw (-1.3,-1.2) node {$a$};

\draw (3.0,-1.4) node {$c$};

\draw (1.0,.5) node {$e$};

\draw[blue, dashed] (3,1) .. controls (6,3) and (6,-3) .. (3,-1);

\end{tikzpicture}

\caption{Wilson networks with loops around non-contractible cycles on the (rigid) torus. Topologically different identifications of endpoints on the second and third graphs yield 2-point blocks in two OPE channels.   } 

\label{12net}
\end{figure}
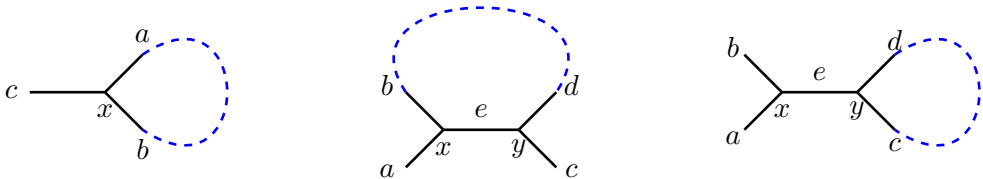

The general idea is that we can build torus blocks from the Wilson networks described in the sphere topology case simply by gluing together any two extra edges modulo $2\pi \tau$ (see Fig. \bref{12net}),  and then identifying the corresponding  representations. Taking a trace in this representation one retains the overall $\sl2$ gauge covariance. More concretely, one takes $(n+2)$-point sphere function \eqref{Phi_s} with $n+2$ boundary states in $\cD_{j_l}$, $l=1,...,n+2$ with any two of them belonging to the same  representation, say $\cD_{j_{1}} \approx \cD_{j_{k}}$ for some $k$. Then, taking a trace over $\cD_{j_{1}}$ naturally singles out  a part of the original $(n+2)-$valent intertwiner involving two Wilson line operators and a number of constituent $3$-valent intertwiners ($k$, for the above choice). By means of the intertwiner invariance property,  the two Wilson operators  can be pulled through the intertwiners to form a single Wilson loop-like operator, schematically,  $\tr_{j_1}\Big(W_{j_1}[0,2\pi\tau] \,\I_{j_1;a,b } \ldots \I_{c;d,j_1} \Big)$. A true Wilson loop is obtained only when one starts from 2-point sphere function and in this case we get the $\sl2$ character (see below), while for higher-point functions an operator under the trace necessarily contains at least one intertwiner.

Let us demonstrate how this works for the trivalent function \eqref{tri2} on the torus $\mathbb{T}^2$ with local coordinates $(w, \bar w)$ giving rise to  a toroidal one-point Wilson network. The Wilson operators here are built using the respective background gravitational connection \eqref{con-glob-13}. We identify any two endpoints of the trivalent graph on Fig. \bref{12net}, which means that  points   $w_1  =  -2\pi  \tau $ and $w_2= 0$ lie on the thermal cycle. Identifying $\cD_a \cong  \cD_b$, then choosing $|a\rangle  = |b\rangle = | j_a,m\rangle$ and summing up over  all basis states in $\cD_a$ (recall that the standard basis is orthonormal)  we find from \eqref{tri2} that  
\be
\label{tri3}
\ba{c}
\dps
\stackrel{\circ}{V}_{a|c}(\tau, {\bf w}) =  \sum_m \Big(\langle j_a,m| \, W_a[-2\pi \tau,0] \,\I_{a; a, c}\, |j_a,m\rangle\Big)\, W_c[0,w] | c\rangle\; 
\\
\\
= \tr_a \Big(W_a[0,2\pi\tau] \,\I_{a; a, c} \Big)W_c[0,w] |c\rangle \;,
\ea
\ee
where by $\stackrel{\circ}{V}$ we denote the resulting toroidal vertex function with some $|c\rangle \in \cD_c$.  If $\cD_c$ is a trivial representation (i.e. $j_c=0$), then the Wilson line operator $W_c = \mathbb{1}_c$ and the intertwiner $\I_{a;a,0} = \mathbb{1}_a$ so that \eqref{tri3} goes to the Wilson loop operator,  
\be
\label{tri4}
\ba{c}
\dps
\stackrel{\circ}{V}_{a|0}(\tau) 
= \tr_a \Big(W_a[0,2\pi \tau]\Big)\;,
\ea
\ee
which is known to be a character of the representation $\cD_a$ \cite{Witten:1988hf}. For non-trivial representations $\cD_c$ we can choose $| c\rangle$ to be a lowest-weight vector in $\cD_c$ and  obtain the expression conjectured in \cite{Kraus:2017ezw}. In Section \bref{sec:gauge} we explicitly check that the expression \eqref{tri3} reproduces the 1-point torus block \eqref{glob_poly}. 

Let us now turn to the two-point toroidal Wilson networks and consider the rightmost graph on Fig. \bref{12net}. Here, the representations labelled by $a,b,c,d$ are associated with endpoints ordered as $w_1,w_2,w_3,w_4$. In terms of the vertex function \eqref{four2} we identify representations $\cD_d \cong \cD_c$ and respective endpoints $w_4 = -2\pi \tau$ and $w_3  = 0$. Now, choosing $|d\rangle  = |c\rangle = |j_c,m\rangle$ and summing up over all $m$ to produce a trace over $\cD_c$, from \eqref{four2} we directly obtain 
\be
\label{four_t}
\stackrel{\circ}{V}_{\hspace{-1mm}{\rm (t)} \, c,e|a,b}(\tau, {\bf w}) = \tr_c\Big( W_c[0,2\pi\tau] \,\I_{c; c, e}\Big)\I_{e; a, b} W_a[0,w_1] W_b[0,w_2]\, | a\rangle \otimes | b\rangle\;.
\ee

The other possible toroidal two-point Wilson network corresponds to the middle graph on Fig. \bref{12net}. We fix  endpoints as $w_4 = -2\pi \tau$ and $w_2 = 0$. Identifying representations $\cD_d \cong \cD_b$ and then  summing  up over states  $|d\rangle  = |b\rangle = |j_b,m\rangle$ we find   
\be
\label{four_s}
\stackrel{\circ}{V}_{\hspace{-1mm}{\rm (s)} \,b,e|a,c}(\tau, {\bf w}) 
=\tr_b\Big(W_b[0,2\pi\tau] \,\I_{b; c, e}\, \I_{e; a, b}\Big)  W_a[0,w_1]\,W_c[0,w_3]  \, |a\rangle \otimes |c\rangle \;.
\ee

Using the crossing equations \eqref{crossing} we see that two-point toroidal vertex functions are related by means of the Wigner $6j$ symbols. In the next sections we check that the vertex functions \eqref{four_s} and \eqref{four_t} with $|a\rangle, |b\rangle$ chosen as lowest-weight vectors  calculate two-point global torus conformal blocks in respectively  $t$-channel and $s$-channel.  Finally, let us note that both functions \eqref{four_t} and \eqref{four_s} are consistently reduced to \eqref{tri3} if one of external spins vanishes. E.g. we can set $j_a=0$ in which case $\I_{e; 0, b}\sim \delta_{eb}\mathbb{1}_e$ \eqref{valent2} and $W_a[0,w_1] =\mathbb{1}_a$. The same is true at $j_b=0$. In other words, two-point vertex functions do reproduce one-point vertex functions provided one of extra spins is zero. The respective torus conformal blocks share the same property.   

\section{Global torus blocks}
\label{sec:one-point}

In this section we  review one-point and two-point torus global blocks and find their explicit form when quasi-primary operators are degenerate, which through the operator-state correspondence are described by finite-dimensional representations of the global conformal algebra.\footnote{\label{fn6}Global blocks are associated with $sl(2,\mathbb{C})$ subalgebra of Virasoro algebra $Vir$ which can be obtained by the \.In\"{o}n\"{u}-Wigner contraction at $1/c \to 0$. Various other limiting blocks can be obtained from $Vir$ by different types of contractions, for details see~\cite{Alkalaev:2016fok} and references therein. Global torus (thermal) blocks in CFT$_d$ and their dual AdS$_{d+1}$ interpretation in terms of the bulk (Witten) geodesic networks were studied in \cite{Alkalaev:2016ptm,Kraus:2017ezw,Alkalaev:2017bzx,Gobeil:2018fzy}.} In what follows  we work in the planar coordinates $(z,\bar z)\in \mathbb{C}$ related to the cylindrical coordinates $(w, \bar w) \in \mathbb{T}^2$ on the torus by the conformal mapping $w = i \log z$.

Prior to describing global blocks let us shortly discuss the origin and relevance of degenerate $sl(2)$ operators. Since $sl(2)\subset Vir$, conformal dimensions of degenerate quasi-primaries can be seen as the large-$c$ limit of the Kac degenerate dimensions $h_{r,s}$ with integer $r,s\geq 1$ \cite{DiFrancesco:1997nk}. Expanding around $c=\infty $ we get 
\be
\label{kac}
h_{r,s} = c\,\frac{1-r^2}{24} - \frac{s-1}{2} -\frac{(1-r)(13r-12s+13)}{24} + \cO(1/c)\;.
\ee      
It follows that in the large-$c$ regime one can distinguish between light ($h \sim \cO(c^0)$) and heavy ($h \sim \cO(c^1)$) degenerate operators. Moreover, those with $h_{1,s}$ are always light,
\be
\label{sl_deg}
h_{1,s} = - \frac{s-1}{2} + \cO(1/c)\;,
\ee    
while heavy operators have $h_{r,s}$ with $r>1$. The paradigmatic example here are the lowest dimension operators: the light operator with $h_{1,2}$ and heavy operator with $h_{2,1}$.\footnote{These are operational in the so-called monodromy method of calculating the large-$c$ (classical) $n$-point conformal blocks via $(n+1)$-point blocks with one light degenerate insertion $h_{1,2}$ \cite{Zamolodchikov1986}. It is intriguing that the monodromy method has a direct AdS dual interpretation as a worldline formulation of particle's configurations \cite{Hijano:2015rla,Alkalaev:2015lca,Banerjee:2016qca,Alkalaev:2016rjl}.} On the other hand, the $sl(2)$ subalgebra with all its representations can be viewed as the Virasoro algebra \.In\"{o}n\"{u}-Wigner contraction (see the footnote \bref{fn6}). Then, the formula \eqref{sl_deg} treats   $1/c$ as a small contraction parameter and  degenerate Virasoro primaries with $h_{1,s}$ in the large-$c$ limit go to the degenerate $sl(2)$ operators corresponding to finite-dimensional non-unitary $sl(2)$ modules with (half-)integer spins $j = -(s-1)/2$, where $s=1,2,3,...\,$.      

From the general physical perspective, degenerate conformal operators constitute spectra of minimal $\cW_N$ models, and, in particular, the Virasoro $Vir=\cW_2$ minimal models relevant in our case.  Moreover, a specific class of minimal coset CFT$_2$ models was conjectured to be dual to $3d$ higher-spin gravity \cite{Gaberdiel:2010pz,Prokushkin:1998bq}. Complementary to the standard 't Hooft limit in such models one may consider the large-$c$ regime limit  \cite{Gaberdiel:2012ku,Perlmutter:2012ds}. Despite that the boundary theory becomes non-unitary (the conformal dimensions are negative, e.g. as discussed above) such a regime is interesting since the bulk gravity is semiclassical according to the Brown-Henneaux relation $G_3 \sim 1/c$ \cite{Brown:1986nw}. Moreover, a gravity dual theory can be formulated as the Chern-Simons theory that brings us back to the study of Wilson lines and conformal blocks, now for such non-unitary finite-dimensional representations \cite{Fitzpatrick:2016mtp,Besken:2017fsj,Hikida:2017ehf,Hikida:2018eih,Hikida:2018dxe}.\footnote{Yet another related direction where non-unitary large-$c$ Virasoro blocks are used is the study of the black hole thermodynamics and information loss in AdS$_3$/CFT$_2$ as a consequence of Virasoro symmetry both in the $c=\infty$ limit and with $1/c$ corrections  \cite{Fitzpatrick:2016ive}.}

\subsection{One-point blocks}

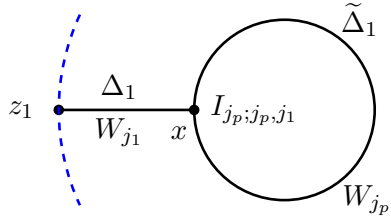
\begin{figure}[H]
\centering

\begin{tikzpicture}[line width=1pt]

\draw[black] (-9,0) circle (1.2cm);
\draw[black] (-10.2,0) -- (-12,0);

\fill[black] (-10.2,0) circle (0.7mm);
\fill[black] (-12,0) circle (0.7mm);

\draw (-8.,1.2) node {$\tDelta_1$};
\draw (-11.2,.3) node {$\Delta_1$};
\draw (-11.2,-.3) node {$W_{j_1}$};
\draw (-7.9,-1.2) node {$W_{j_p}$};

\draw (-12.5,0) node {$z_1$};
\draw (-10.4,-0.3) node {$x$};
\draw (-9.4,-0.) node {$I_{j_p;j_p,j_1}$};

\centerarc[dashed,blue](-9,0)(155:205:3cm);

\end{tikzpicture}
\caption{One-point conformal block on a torus. $\Delta$ and $\tDelta$ are external and intermediate conformal dimensions. The same graph shows the Wilson network with Wilson line $W_{j_1}$ and loop $W_{j_p}$, along with the intertwiner $I_{j_p;j_p,j_1}$ in the bulk point $x$. Dashed line is the boundary of the thermal AdS$_3$.  }
\label{1-point-fig}
\end{figure}

The global one-point block in the torus CFT$_2$ is defined as the holomorphic contribution to the one-point correlation function of a given (quasi-)primary operator,  
\be
\label{1pt}
\langle \cO_{\Delta, \bar \Delta}(z,\bar z)\rangle  = \tr\left(q^{L_0} \bar q^{\bar L_0} \cO_{\Delta, \bar \Delta}(z,\bar z) \right)  = \sum_{^{\tilde \Delta, \bar{\tilde \Delta}}}  C^{^{\Delta \bar \Delta}}_{^{\tilde \Delta \bar{\tilde \Delta}}}\;\cF_{^{\tilde\Delta,\Delta}}(z,q)\,\cF_{^{\bar{\tilde \Delta}, \bar \Delta }}(\bar z,\bar q)\;,
\ee
where the trace $\tr$ is taken over the space of states, and the right-hand side defines the OPE expansion into (anti)holomorphic torus blocks, the expansion coefficients are the 3-point structure constants with two dimensions identified that corresponds to creating a  loop (see  Fig. \bref{1-point-fig}). The parameter $q = \exp{2\pi i \tau}$, where $\tau\in \mathbb{H}$ is the torus modulus, (holomorphic) conformal dimensions $\Delta, \tilde \Delta$ parameterize the external (quasi-)primary operator and the OPE channel, respectively. The convenient representation of the torus block is given by using the hypergeometric function~\cite{Hadasz:2009db}
\be
\label{glob1pt}
\begin{aligned}
&\cF_{^{\tilde\Delta,\Delta}}(q) = \frac{\;\;q^{ \tilde \Delta}}{1-q} \,\;{}_2 F_{1}(\Delta, 1-\Delta, 2\tilde \Delta\, |\, \frac{q}{q-1})  =
\\
& = 1+\Big[1+\frac{(\Delta - 1)\Delta}{2\tilde \Delta}\Big] q + \Big[1+\frac{(\Delta -1) \Delta }{2 \tilde\Delta }+\frac{(\Delta -1)
 \Delta  (\Delta ^2-\Delta +4 \tilde\Delta)}{4 \tilde \Delta  (2 \tilde\Delta +1)}\Big]q^2+ ... \;.
\end{aligned}
\ee
The one-point block function is $z$-independent due to the global $u(1)$ translational invariance of torus CFT$_2$. Note that at $\Delta=0$ the one-point function becomes the $\sl2$ character 
\be
\label{sl2_char}
\cF_{^{\tilde\Delta, 0}}(q) = \frac{\;\;q^{ \tilde \Delta}}{1-q} = q^{ \tilde \Delta}\left(1+q+q^2+q^3+\ldots\right)\,,
\ee
showing that for generic dimension $\tilde\Delta$ there is one state on each level of the corresponding Verma module. The block function \eqref{glob1pt} can be shown to have poles at 
$\tilde \Delta = -n/2$, where $n\in \mathbb{N}_0\,$ which is most manifest when  the block is represented as the Legendre function (see Appendix \bref{appC}).

In general, conformal dimensions $\Delta,\tDelta$ are assumed to be arbitrary. The corresponding  operators are related to (non)-unitary infinite-dimensional $\slC2$ representations. For integer negative dimensions\footnote{See Appendix \bref{sec:sl2}. In this paper we consider only bosonic (integer spin) representations.}
\be
\label{deg_del}
\Delta=-j_1\quad\text{and}\quad \tilde\Delta=-j_p\;,
\qquad
j_1, j_p \in \mathbb{N}_0\;,
\ee
these representations contain singular submodules on level $2j$ with the conformal dimension $\Delta' = -\Delta+1$  so that one may consider quotient representations $\cD_j$ which are finite-dimensional non-unitary spin-$j$ representations of dimension $2j+1$. 

Note that the degenerate dimension $\tDelta$ of the loop channel \eqref{deg_del} defines poles of the block function.  It follows that in order to find  torus blocks for $\cD_j$ one may proceed in two equivalent ways. The first one is to use the BPZ procedure \cite{Belavin:1984vu}  to impose the singular vector decoupling condition on correlation functions. E.g., for the zero-point blocks which are characters \eqref{sl2_char} the BPZ condition  is solved by subtracting the singular submodule character $\cF_{^{\tilde\Delta',0}}$ from the original character $\cF_{^{\tilde\Delta,0}}$  to obtain the well-known expression for finite-dimensional $\slC2$ character
\be
\label{fin_char}
\chi_p \equiv \cF_{^{\tilde\Delta,0}}(q) - \cF_{^{\tilde\Delta',0}}(q)
= q^{-j_p}\left(1 + q  + q^2 +... + q^{2j_p}\right) 
= q^{-j_p} \frac{\;\;1-q^{j_p+1}}{1-q}\;.
\ee
Similarly, one may formulate and solve the respective BPZ condition for one-point blocks.

The alternative way is to define the torus block as (anti)holomorphic constituents of the correlation functions built by evaluating the trace already in the finite-dimensional quotient modules. In this case, the holomorphic one-point block is simply 
\be
\label{fin_1pt}
\cF_{{j_p, j_1}}(q)  = \tr_{j_p} \left(q^{L_0}\cO_{j_1}\right)\;,
\ee
where the quasi-primary operator $\cO_{j_1}$ corresponds to $\cD_{j_1}$, and the trace is taken over $\cD_{j_p}$, cf. \eqref{1pt}.  It becomes the order-$2j_p$ polynomial in the modular parameter,
\be
\label{glob_poly}
\cF_{{j_p, j_1}}(q)=q^{-j_p}\left(f_{0} + q  f_{1} + q^2 f_{2} +... + q^{2j_p} f_{2j_p}\right)\;,
\qquad
\ee
where the coefficients  are given by 
\be
\label{1ptcoefGLOBAL}
f_{n}  = \, _3F_2(-j_1,j_1+1,-n;1,-2 j_p;1)= 
\sum_{m=0}^n
\,\frac{(-)^m n!}{(n-m)!(m!)^2}\,\frac{(j_1+m)_{2m}}{(2j_p)_m}
\ee
and $(x)_n = x(x-1)... (x-n+1)$ is the falling factorial. At $j_1 =0$ we reproduce \eqref{fin_char}. Note that imposing the BPZ condition enforces  the conformal dimensions to satisfy the fusion rule  
\be
\label{restrict}
0 \leq j_1 \leq 2j_p\;.
\ee    

\subsection{Two-point blocks }

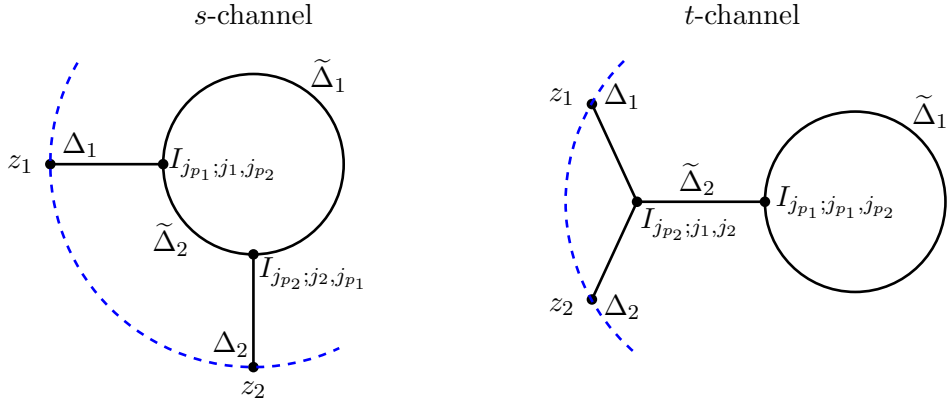
\begin{figure}[H]
\centering

 \begin{tikzpicture}[line width=1pt]
 
 \def\x{8}
 \def\y{-.5}


\draw[black] (-9,0) circle (1.2cm);
\draw[black] (-10.2,0) -- (-11.7,0);
\draw[black] (-9,-1.2) -- (-9,-2.7);

\draw (-10.1,-1.) node {$\tDelta_2$};
\draw (-8.,1.18) node {$\tDelta_1$};
\draw (-8.6-0.7,-2.4) node {$\Delta_2$};
\draw (-11.3,.25) node {$\Delta_1$};

\draw[black] (-9+\x,0+\y) circle (1.2cm);
\draw[black] (-10.2+\x,0+\y) -- (-11.9+\x,0+\y);
\draw[black] (-12.5+\x,1.3+\y) -- (-11.9+\x,0+\y);
\draw[black] (-12.5+\x,-1.3+\y) -- (-11.9+\x,0+\y);

\draw (-8+\x,1.2+\y) node {$\tDelta_1$};
\draw (-11.1+\x,.3+\y) node {$\tDelta_2$};
\draw (-12.1+\x,1.4+\y) node {$\Delta_1$};
\draw (-12.1+\x,-1.4+\y) node {$\Delta_2$};

\draw (-9,2) node {$s\text{-channel}$};
\draw (-2.5,2) node {$t\text{-channel}$};

\draw (-12.1,0) node {$z_1$};
\draw (-9.,-3) node {$z_2$};

\fill[black] (-10.2,0) circle (0.7mm);
\fill[black] (-11.7,0) circle (0.7mm);
\fill[black] (-9,-2.7) circle (0.7mm);
\fill[black] (-9,-1.2) circle (0.7mm);

\fill[black] (-10.2+\x,0+\y)  circle (0.7mm);
\fill[black]  (-11.9+\x,0+\y)  circle (0.7mm);
\fill[black] (-12.5+\x,1.3+\y) circle (0.7mm);
\fill[black] (-12.5+\x,-1.3+\y) circle (0.7mm);

\draw (-12.1+\x-.8,1.4+\y) node {$z_1$};
\draw (-12.1+\x-.8,-1.4+\y) node {$z_2$};

\draw (-9.4,-0.) node {$I_{j_{p_1};j_1,j_{p_2}}$};
\draw (-8.2,-1.45) node {$I_{j_{p_2};j_2,j_{p_1}}$};
\centerarc[dashed,blue](-9,0)(150:295:2.7cm);

\draw (-9.25+\x,-0.+\y) node {$I_{j_{p_1};j_{p_1},j_{p_2}}$};
\draw (-11.2+\x,-0.+\y-.3) node {$I_{j_{p_2};j_{1},j_{2}}$};
\centerarc[dashed,blue](-10.2+\x,0+\y)(135:230:2.65cm);

\end{tikzpicture}
\caption{Two-point conformal blocks in $s$-channel and $t$-channel. The same graphs show the Wilson networks, with dashed lines representing the boundary of the thermal AdS$_3$. }
\label{2-point-fig}
\end{figure}

The global two-point torus correlation functions can be expanded in two OPE channels that below  are referred  to  as $s$-channel and $t$-channel, see  Fig. \bref{2-point-fig}. 
 
\paragraph{$s$-channel.} The two-point $s$-channel global block   is given by \cite{Alkalaev:2017bzx}:   
\be
\label{glob-s}
\begin{aligned}
&\cF_s^{^{\Delta_{1,2}, \tilde \Delta_{1,2}}}(q, z_{1,2}) =\\
&\quad=z_1^{-\Delta_1+\tilde\Delta_1-\tilde\Delta_2} z_2^{-\Delta_2-\tilde\Delta_1+\tilde\Delta_2} \sum_{m,n=0}^\infty \frac{\tau_{n,m}(\tilde\Delta_1, \Delta_1, \tilde\Delta_2)\tau_{m,n}(\tilde\Delta_2, \Delta_2, \tilde \Delta_1)}{m!\, n!\,(2\tilde\Delta_1)_n (2\tilde\Delta_2)_m}\, q^{\tilde \Delta_1+n}\, \left(\frac{z_1}{z_2}\right)^{n-m}\;,
\end{aligned}
\ee
where the coefficients $\tau_{m,n} = \tau_{m,n}(\Delta_i, \Delta_j, \Delta_k)$ defining  the $sl(2)$ 3-point function of  a primary operator $\Delta_j$ and  descendant operators  $\Delta_{i,k}$ on the levels $n,m$ are~\cite{Alkalaev:2015fbw}
\be
\label{A-tau}
\tau_{n,m}(\Delta_{i,j,k}) = \sum_{p = 0}^{\min[n,m]} 
\left(\begin{array}{c}
 n \\
 p \\
\end{array}
\right)
 (2\Delta_k +m-1)^{(p)} m^{(p)}
(\Delta_k+\Delta_j - \Delta_i)_{m-p}(\Delta_i + \Delta_j -\Delta_k+p-m)_{n-p}\;,
\ee
where $(x)_l = x(x+1)...(x+l-1)$ and $(x)^{(l)} = x(x-1)... (x-l+1)$ are raising (falling) factorials. 

The conformal dimensions of the degenerate (external and internal) quasi-primary operators read   
\be
\label{degen}
\Delta_a=-j_a\quad\text{and}\quad \tilde\Delta_b=-j_{p_b}\;,
\qquad
a,b=1,2\;,
\qquad
j_a, j_{p_b} \in \mathbb{N}_0\;.
\ee
These values of the intermediate channel dimensions  define poles of the block function. It follows that 
in order to use~\eqref{glob-s} for the degenerate operators  the summation is to be restricted to the region $n<  -2\tilde\Delta_1+1$ and $m<  -2\tilde\Delta_2+1$ which is an implementation of the BPZ decoupling condition.

\paragraph{$t$-channel.}  The two-point global block in the $t$-channel can be calculated either by solving the $sl(2, \mathbb{C})$ Casimir equation \cite{Kraus:2017ezw} or by summing over 3-point functions of three descendant operators  \cite{Alkalaev:2017bzx}: 
\be
\label{glob-t}
\begin{aligned}
&\cF_{\, t}^{^{\Delta_{1,2}, \tilde \Delta_{1,2}}}(q, z_{1,2}) =\\
&\quad=(z_1-z_2)^{\tilde\Delta_2-\Delta_1-\Delta_2} z_2^{-\tilde\Delta_2} 
\sum_{m,n=0}^\infty \frac{\sigma_{m}(\Delta_{1}, \Delta_{2}, \tilde\Delta_2)\tau_{n,n}(\tilde\Delta_1, \tilde\Delta_2, \tilde\Delta_1)}{m!\,n!\,(2\tilde\Delta_1)_n(2\tilde\Delta_2)_m} \, q^{\tilde \Delta_1+n} \, \left(\frac{z_1-z_2}{z_2}\right)^m,
\end{aligned}
\ee
where the $\tau$-function is given by \eqref{A-tau} and $\sigma_{m}(\Delta_{1}, \Delta_{2}, \tilde\Delta_2) = (-)^m (\tilde\Delta_2+\Delta_1-\Delta_2)_{m}(\tilde \Delta_2)_{m}$. The $t$-channel block for the degenerate quasi-primary operators is obtained by applying arguments given around \eqref{degen}.

\section{One-point toroidal Wilson networks} 
\label{sec:gauge}

In this section we explicitly calculate the one-point toroidal vertex function \eqref{tri3} in the thermal AdS$_3$ \eqref{con-glob-13} which  is equal to the one-point block function \eqref{glob_poly}--\eqref{1ptcoefGLOBAL}. 

\subsection{Diagonal gauge} 
\label{sec:diagonal}

Let us choose the boundary state as the lowest-weight vector in the respective spin-$j_a$ representation $\cD_a$ (see Appendix \bref{sec:sl2})
\be
\label{LW_i}
|a\rangle =|\mathbb{lw}\rangle_a\in \cD_a\;:
\qquad 
J_{1}|\mathbb{lw}\rangle_a = 0\;,
\qquad
J_{0}|\mathbb{lw}\rangle_a = -j_a |\mathbb{lw}\rangle_a\;, 
\ee 
along with the Wilson line operator in Euclidean thermal AdS$_3$ space defined by the gravitational  connection \eqref{con-glob-13}. Performing a  gauge transformation, $\a=U \,\tilde{\a} \, U^{-1} +UdU^{-1}$, where the $SL(2,\mathbb{R})$ gauge group element is $U = \exp{\frac{i}{2}J_{-1}}\exp{-iJ_1}\exp{-\frac{i \pi}{2}J_{0}}$, the chiral  connection~\eqref{con-glob-13} can be cast into the {\it diagonal} form
$\tilde{\a}= -i J_0\, dw$ as compared to the original {\it off-diagonal} form.\footnote{See \cite{Castro:2011fm} for more about the diagonal gauge in the $sl(N, \mathbb{R})$ higher-spin gravity case.}  In the diagonal gauge, the chiral Wilson line operators \eqref{chiral_wilson} take the following general  form 
\be
W_a[x_1,x_2] = \exp{(-ix_{12}J_0)}\;.
\ee
The Wilson loop traversing the thermal circle can now be given in terms of the modular parameter as 
\be
\label{Wilson_loop}
W_{a}[0,2\pi \tau]=\exp{(2\pi i \tau J_0)} \equiv q^{J_0}\;, 
\qquad
q = \exp{2\pi i\tau}\;,
\ee
where $J_0$ is taken in the representation $\cD_a$. Due to the intertwiner transformation property, using the Wilson operators in  the diagonal gauge does not change toroidal vertex functions of Section \bref{sec:toroidal} except for that the boundary lowest-weight vectors \eqref{LW_i} are to be transformed accordingly, 
\be
\label{U_trans}
|\mathbb{lw}\rangle_a \to |\hat{\mathbb{lw}\,}\rangle_a  = U^{-1}_a |\mathbb{lw}\rangle_a \in \cD_a\;,
\ee    
with the  gauge group (constant) element $U_a$  given above and evaluated in the representation $\cD_a$. In the diagonal gauge, the conformal invariance of the toroidal vertex functions is guaranteed by the property \eqref{conf_trans} which is now given by the identity matrix $C_m{}^n = \delta_m^n$ and the holomorphic conformal algebra generators $\cL_n = -\exp{(inw)}(i \partial_w-n\Delta)$ in the cylindrical coordinates on $\mathbb{T}^2$.

The transformed boundary vector \eqref{U_trans} obeys the following linear relations \cite{Kraus:2017ezw}
\be
\begin{aligned}
\label{special-state-1}
\left(J_1+J_{-1}-2J_0\right)  |\hat{\mathbb{lw}\,}\rangle =0\;,\\
\left(J_1 - J_{-1} +2j\right)  |\hat{\mathbb{lw}\,}\rangle=0\;,
\end{aligned}
\ee
which are the two transformed lowest-weight conditions \eqref{LW_i}.  Representing $|\hat{\mathbb{lw}\,}\rangle \in \cD_j$ in the standard  basis as
\be
\label{trans1}
|\hat{\mathbb{lw}\,}\rangle=\beta_{j,j}|j,j\rangle+\beta_{j,j-1} |j,j-1\rangle+ \cdots+ \beta_{j,-j} |j,-j\rangle\;,
\ee 
acting on this state with $J_1$ or $J_{-1}$ and then using the defining relations \eqref{special-state-1} we obtain the recurrence 
relations
\be
\label{rec_vec}
\begin{aligned}
&\beta_{j,k} (h_j+k)=M_-(j,k+1) \beta_{j,k+1}\;,\\
&M_+(j,k-1) \beta_{j,k-1}=(k-h_j)\beta_{j,k}\;,
\end{aligned}
\ee
where $k=-j,-j+1, ...,j-1, j$ and  $M_{\pm}(j,k)$ are defined in~\eqref{standard}. Fixing the overall normalization as  $\beta_{j,j} = 1$ and solving~\eqref{rec_vec}  we find
\be
\label{trans2}
\beta_{j,k}=- \prod _{r=0}^{j-k-1} \frac{M_-(j,j-r)}{r+1}
=
\prod _{r=0}^{j-k-1} \frac{2j-r}{M_+(j,j-r-1)}\;.
\ee

In what follows we will need the transformed boundary state in the fundamental (two-dimensional) representation  $\cD_{\half}$ which is read off from \eqref{trans1}, \eqref{trans2} as
\be
\label{fund-state-tilde}
|\hat{\mathbb{lw}\,}\rangle=|\half,\half\rangle+|\half,-\half\rangle\;.
\ee

\subsection{Wigner \3j symbol representation} 
\label{sec:wigner}

Let us consider the one-point toroidal Wilson network in the diagonal gauge. To this end, using the translation invariance we set $w=0$ so that $W_c = \mathbb{1}_c$, then insert the resolutions of identities \eqref{resolve} that allows representing the vertex function \eqref{tri3} in the form  
\be
\label{tri_m}
\ba{c}
\dps
\stackrel{\circ}{V}_{a|c}(\tau) =  \sum_{m,n = -j_a}^{j_a} \sum_{l=-j_c}^{j_c} \Big(\langle j_a,m| \, W_a[0,2\pi \tau] \,|j_a,n\rangle\Big) \Big(\langle j_a,n|\I_{a; a, c}\, |j_a,m\rangle\otimes |j_c, l\rangle\Big)\langle j_c, l|\hat{\mathbb{lw}\,}\rangle_c\;,
\ea
\ee 
where the Wilson  operator  is given by \eqref{Wilson_loop} and $|\hat{\mathbb{lw}\,}\rangle_c$ is the transformed boundary  vector \eqref{U_trans}. The first factor in \eqref{tri_m} is the Wigner D-matrix for the $SU(1,1)$ group element \eqref{Wilson_loop},  
\be\label{D-matrix}
D^{(j_a)}{}^m{}_n \equiv \langle j_a,m| \, q^{J_0} \,|j_a,n\rangle = \delta^{m}_n \,q^n\;,
\qquad
m,n = -j_a, -j_a+1, ... , j_a\;,
\ee      
where the last equality is obtained by using the standard basis \eqref{standard}.\footnote{Note that taking the trace of the D-matrix \eqref{D-matrix} we directly obtain the $su(1,1)$ character \eqref{fin_char}.} The last factor is the $l$-th coordinate of the transformed boundary vector in the standard basis,
\be
V_{(j_c)}^l \equiv  \langle j_c,l|\hat{\mathbb{lw}\,}\rangle_c\;,
\qquad
l = -j_c, -j_c+1, ... , j_c\;,
\ee 
It is given by \eqref{trans1}, \eqref{trans2}. Finally, the second factor is the matrix element of the intertwiner which is related to the Wigner \3j symbol by \eqref{tri3-1}. Gathering all matrix elements together we obtain 
\be
\label{tri_mat}
\ba{c}
\dps
\stackrel{\circ}{V}_{a|c}(q) =  \sum_{m,n = -j_a}^{j_a} \sum_{l=-j_c}^{j_c} 
D^{(j_a)}{}^m{}_n [I_{a;a,c}]^n{}_{ml} V_{(j_c)}^l 
= 
(-)^{j_a}V_{(j_c)}^0\sum_{m = -j_a}^{j_a}  (-)^m q^m   
\begin{pmatrix}
j_c & j_a &\, j_a \\
0 & m & \,-m 
\end{pmatrix}\;,
\ea
\ee
where when obtaining the second equality  we used: (1) relations \eqref{mat_int}, \eqref{tri3-1}, (2)  the \3j symbol $[W_{a,b,c}]_{mkn}$  property $m+n+k=0$, (3) the D-matrix $D^{(j_a)}{}^m{}_n$ is diagonal, $m-n=0$. The last two properties allow  reducing  three sums to one. Also, it follows that  only the middle (the magnetic number =0) component of the boundary transformed vector contributes by giving an overall factor.

Now, we adjust representations $\cD_a\approx \cD_{j_p}$ and $\cD_c\approx \cD_{j_1}$ to the loop and the external leg. In this case, the sum in the right-hand side of \eqref{tri_mat} can be cast into the form  
\be
\label{tri_mat2}
\sum_{m=-j_p}^{j_p} (-)^m
\begin{pmatrix}
j_1 & j_p &\, j_p \\
0 & m & \,-m 
\end{pmatrix}\,q^m =
q^{-j_p}\sum_{n=0}^{2j_p} (-)^{n-j_p}
\begin{pmatrix}
j_1 & j_p & j_p \\
0 & n-j_p & j_p-n 
\end{pmatrix}\,q^n\;,
\ee
which is obtained by changing $n=m+j_p$. 
 On the other hand, the expansion coefficients of the one-point block~\eqref{glob_poly} are given by the hypergeometric function \eqref{1ptcoefGLOBAL}. Hence, in order to verify the representation~\eqref{tri_mat} we need to check the identity  
\be
\label{two-hyper-geoms}
 \, _3F_2(-j_1,j_1+1,-n;1,-2 j_p;1)=\varkappa\, (-)^{n-j_p}
 \begin{pmatrix}
j_1 & j_p & j_p \\
0 & n-j_p & j_p-n 
\end{pmatrix}\,,
\ee 
where $\varkappa$ is an  $n$-independent factor. This relation holds for  
\be
\varkappa=(-)^{j_p}
\begin{pmatrix}
j_1 & j_p &\, j_p \\
0 & j_p &\, -j_p 
\end{pmatrix}^{-1}=
(-)^{j_p}\, \frac{(2 j_p+1)\sqrt{ \Gamma (1+2 j_p-j_1) \Gamma (2+2 j_p+j_1)}}{\Gamma (2 j_p+2)}\,.
\ee
To see this we use the  explicit representation for the Wigner \3j symbol~\cite{varshalovich}
\be
\label{3j-explicit}
\begin{aligned}
&\begin{pmatrix}
j_1 & j_2 & j_3 \\
m_1 & m_2 & m_3 
\end{pmatrix} =\delta _{m_1+m_2+m_3,0}\,
\frac{\sqrt{\left(j_3-j_1+j_2\right)!} \sqrt{\left(-j_3+j_1+j_2\right)!} \sqrt{\left(j_3+j_1+j_2+1\right)!}}{\Gamma \left(j_3+j_1+j_2+2\right)\sqrt{\left(j_3+j_1-j_2\right)!}}\times\\
&\times\frac{\sqrt{(j_3-m_3)!} \sqrt{\left(j_1-m_1\right)!}}{\sqrt{(j_3+m_3)!} \sqrt{\left(j_1+m_1\right)!} \sqrt{\left(j_2-m_2\right)!} \sqrt{\left(j_2+m_2\right)!}}\,
\frac{(-)^{j_1+m_2-m_3}(2 j_3)!  \left(j_3+j_2+m_1\right)!}{\left(j_3-j_1+j_2\right)!  (j_3-m_3)!}\\
&\times \, _3F_2\left(-j_3+m_3,-j_3-j_1-j_2-1,-j_3+j_1-j_2;-2 j_3,-j_3-j_2-m_1;1\right)\;,
\end{aligned}
\ee
which gives for the right-hand side of~\eqref{two-hyper-geoms}
\be
\varkappa \, (-)^{n-j_p}
\begin{pmatrix}
j_1 & j_p & j_p \\
0 & n-j_p & j_p-n 
\end{pmatrix} =
\frac{(2 j_p)! (-1)^{n-j_1} \, _3F_2(-j_1-2 j_p-1,j_1-2 j_p,-n;-2 j_p,-2 j_p;1)}{n! (2 j_p-n)!}\;.
\ee
The right-hand side here can be transformed by making use of the (Euler-type) transformation for the generalized hypergeometric function (see e.g.~\cite{prudnikov1986integrals}),
\be\label{Euler-transform}
\frac{\Gamma (d) \Gamma (-a-b-c+d+e) \, _3F_2(e-a,e-b,c;-a-b+d+e,e;1)}{\Gamma (d-c) \Gamma (-a-b+d+e)}= \, _3F_2(a,b,c;d,e;1)\;,
\ee
so that we obtain the equality~\eqref{two-hyper-geoms} that proves the representation~\eqref{tri_mat} for the one-point torus block~\eqref{glob_poly}.

\subsection{Symmetric tensor product representation }
\label{sec:one_point}

Yet another possible realization of the intertwiners is given when finite-dimensional $\sl2$ representations $\cD_j$ are realized as components of the symmetric tensor products of fundamental (spinor) representation $\cD_{\half}$ (for notation and conventions, see Appendix \bref{sec:multi}). This multispinor technique was used in \cite{Besken:2016ooo} to calculate three-point and four-point sphere blocks. In what follows we explicitly calculate the one-point torus block for degenerate operators using the toroidal vertex function \eqref{tri3} realized via multispinors. In particular,  this realization of  the Wilson network formulation  brings to light an interesting decomposition of the torus block function in terms of $\sl2$ characters (see Section \bref{sec:character}).

We start with the toroidal one-point vertex function given in the form~\eqref{tri_m} or \eqref{tri_mat}, which is a product of three matrix elements which are now to be  calculated using  the multispinor approach.

\paragraph{(I)} \hspace{-3mm} The first matrix element in \eqref{tri_m} is the Wigner $D$-matrix, 
\be
\label{M1}
D_{\alpha_1\cdots\alpha_{\lambda_a}}^{\beta_1\cdots\,\beta_{\lambda_a}} 
= \frac{1}{\lambda_a!} \,D_{(\alpha_1}^{\beta_1}\cdots D_{\alpha_{\lambda_a})}^{\beta_{\lambda_a}} \;,
\ee
where $\lambda_a = 2j_a$ and  $D_\alpha^\beta$ is the  Wigner $D$-matrix  of the Wilson line wrapping the thermal cycle in the fundamental representation, 
\be
\label{Q-tilde-a}
D_\alpha^\beta=\langle e_\alpha|q^{J_0}|e_\beta\rangle=\left(
\begin{array}{cc}
 q^{\half} & 0 \\
 0 & q^{-\half} \\
\end{array}
\right)\;,
\ee  
where $|e_\alpha\rangle = |\half, (-)^{1+\alpha}\half\rangle$ denote the standard basis elements.\footnote{Note that taking the trace of the D-matrix \eqref{M1} one can obtain the $su(1,1)$ character \eqref{fin_char} \cite{Kraus:2017ezw}.}
   
\paragraph{(II)} \hspace{-3mm} The third matrix element in \eqref{tri_m} is coordinates of the transformed boundary state $|c\rangle = |\hat{\mathbb{lw}\,}\rangle_c$ defined by \eqref{special-state-1}. Then, using the product formulas for spinors and representing $|\hat{\mathbb{lw}\,}\rangle_c$ as a product of elements \eqref{fund-state-tilde} we find   
\be
\label{M2}
\langle j_c,l|\hat{\mathbb{lw}\,}\rangle_c \sim V_{\gamma_1\cdots\gamma_{\lambda_c}}=V_{\gamma_1}\cdots V_{\gamma_{\lambda_c}}\;,
\ee   
where a spinor $V_\gamma$ is now coordinates of the transformed boundary vector \eqref{fund-state-tilde} in the fundamental representation,
\be
\label{V-tilde-a} 
V_\gamma=\langle e_\gamma|\Big(|e_1\rangle+|e_2\rangle\Big)=\delta_{\gamma,1}+\delta_{\gamma,2}\;.
\ee

\paragraph{(III)} \hspace{-3mm} The second matrix element  in \eqref{tri_m} is the intertwiner which in the spinor form is just the projector \eqref{proj} with $\lambda_1 = \lambda_3 = \lambda_a$ and $\lambda_2 = \lambda_c$, so that from \eqref{k} we have $k = \lambda_c/2$.

\vspace{3mm}

Gathering all matrix elements together we assemble the following Wilson network matrix element 
\be
\ba{l}
\dps
W^{\rho_1 ... \rho_{\lambda_a}}_{\gamma_1 ... \gamma_{\lambda_a}}=
\\
\\
\dps
\hspace{20mm}= \epsilon^{\alpha_{1}\beta_1}\cdots \epsilon^{\alpha_{\frac{\lambda_c}{2}}\beta_{\frac{\lambda_c}{2}}}\; 
D^{\rho_1 \ldots \rho_{\frac{\lambda_1}{2}}  \rho_{\frac{\lambda_c}{2}+1} \ldots \rho_{\lambda_a}}_{\alpha_1 \ldots \alpha_{\frac{\lambda_c}{2}} (\gamma_{1} \ldots \gamma_{\lambda_a-\frac{\lambda_1}{2}}}
\;V_{\gamma_{\lambda_a-\frac{\lambda_1}{2}+1}\ldots \gamma_{\lambda_a}) \beta_1 \ldots \beta_{\frac{\lambda_c}{2}}}\;,
\ea
\ee
so that the vertex function \eqref{tri_m} is given by 
\be
\stackrel{\circ}{V}_{a|c} = W_{\gamma_1\cdots\gamma_{\lambda_a}}^{\gamma_1\cdots\gamma_{\lambda_a}} \;.
\ee

In the rest of this section we show that the vertex function calculates   the one-point  torus block with degenerate operators as  
\be\label{one-point-tensor}
\cF_{{j_p, j_1}}(q) = W_{\gamma_1\cdots\gamma_{2j_p}}^{\gamma_1\cdots\gamma_{2j_p}}\;.
\ee
Substituting  \eqref{M1} and \eqref{M2} into \eqref{one-point-tensor} we obtain 
\be
\label{Fpk}
\ba{l}
\dps
\cF_{{j_p, j_1}}(q) = \frac{1}{(2j_p)!}\,\epsilon^{\alpha_1 \beta_1}\, \cdots\, \epsilon^{\alpha_{j_1} \beta_{j_1}}
\,
D^{\gamma_1}_{\alpha_1}\, \cdots\,  D^{\gamma_{j_1}}_{\alpha_{j_1}}
\, 
V_{\beta_1} \, \cdots\,  V_{\beta_{j_1}}\, \times
\\
\\
\dps
\hspace{60mm}\times\, D^{\gamma_{{j_1}+1}}_{(\gamma_1} \, \cdots\,  D^{\gamma_{2j_p}}_{\gamma_{{2j_p}-{j_1}}}
\,
 V_{\gamma_{{2j_p}-{j_1}+1}} \, \cdots\,  V_{\gamma_{2j_p})}\,. 
\ea
\ee    

In order to calculate \eqref{Fpk} we parameterize $2j_p = p$ and $j_1 = k$ along with the fusion condition $k\leq p$ \eqref{restrict}. This expression can be simplified by introducing new spinor and scalar  functions, 
\be
E^\gamma = \epsilon^{\alpha\beta} D_\alpha^\gamma V_\beta = (q^{\half}, -q^{-\half})
\ee
and
\be
\label{mathCn}
\mathbb{C}_n = E^{\gamma}\,\left( D^{\alpha}_{\beta_1}D^{\beta_1}_{\beta_2} D^{\beta_2}_{\beta_3} \cdots D^{\beta_{n-1}}_{\gamma}\right)\, V_\alpha \equiv
E^{\gamma}\,\left(D^{n}\right)^\alpha_\gamma\, V_\alpha = q^{\frac{n+1}{2}} - q^{-\frac{n+1}{2}}\;,
\ee
which are calculated using the definitions \eqref{Q-tilde-a} and \eqref{V-tilde-a}.
Then, \eqref{Fpk} can be cast into the form
\be
\label{Fpk1}
\cF_{{j_p, j_1}}(q) \equiv F_{_{p,k}} = \frac{1}{p!}\,E^{\gamma_1} \cdots E^{\gamma_k}\, 
D^{\gamma_{k+1}}_{(\gamma_1} \, \cdots\,  D^{\gamma_p}_{\gamma_{p-k}}
\,
 V_{\gamma_{p-k+1}} \, \cdots\,  V_{\gamma_p)}\;.
\ee    

Let us introduce the following  matrix element, 
\be
\label{Inm}
[\mathbb{T}^{(n,m)}]^{\alpha_1 ... \alpha_m}_{\beta_1 ... \beta_m} =\frac{1}{n!} \,D^{\alpha_1}_{(\beta_1} \cdots D^{\alpha_m}_{\beta_m} D^{\gamma_{m+1}}_{\gamma_{m+1}} \cdots D^{\gamma_{n}}_{\gamma_{n})} 
\ee
at $m=0,...,n$. E.g. at $m=0$ we get the character $[\mathbb{T}^{(n,0)}] \equiv \chi_n$ of the spin-$n/2$ representations  \cite{Kraus:2017ezw}, while at higher $m$ these elements serve as building blocks of the matrix element \eqref{Fpk1}.  To calculate \eqref{Inm} it is convenient to classify all contractions in terms of cycles of the symmetric group $S_n$ acting on the lower indices. Noting that a length-$s$ cycle is given by $s$-th power of the Wigner D-matrix $D^\alpha_\beta$ we find 
\be
\label{square}
[\mathbb{T}^{(n,m)}]^{\alpha_1 ... \alpha_m}_{\beta_1 ... \beta_m} = \frac{1}{n!}\; \sum_{m \leq s_1 + ... +s_m \leq n} b_{s_1, ... , s_m} (D^{s_1})^{\alpha_1}_{(\beta_1} \cdots (D^{s_m})^{\alpha_m}_{\beta_m)}\, [\mathbb{T}^{(n - s_1- ... -s_m,0)}]\;,
\ee
with the coefficients  
\be
\label{5.33}
b_{s_1, ... , s_m} =
(n - s_1- ... -s_m)!\,\prod_{i=1}^m  A_{n-m-s_1-...-s_{i-1}+i-1}^{s_i-1} = (n-m)!\;,
\ee
where
\be
A_s^t = \frac{s!}{(s-t)!}
\ee
denotes the number of partial permutations  (sequences without repetitions). In \eqref{5.33} the first  factorial corresponds to the number of terms in $[\mathbb{T}^{(n - s_1- ... -s_m,0)}]$, while each factor in the product counts a number of independent cycles of length $s_i$ in the original symmetrization of $n$ indices. Remarkably, the result does not depend on $s_i$. 

Using the matrix elements \eqref{Inm} the original expression \eqref{Fpk1} can be represented as
\be
F_{p,k} 
=\sum_{l=0}^{k} a_{l}\, (\mathbb{C}_0)^{l}  
\,E^{\beta_1} \cdots E^{\beta_{k-l}} 
\;[\mathbb{T}^{(p-k,k-l)}]^{\alpha_1\cdots \alpha_{k-l}}_{\beta_1 \cdots \beta_{k-l}}
\;V_{\alpha_1} \cdots V_{\alpha_{k-l}}\;,
\ee 
where coefficients are given by the triangle sequence  
\be
\label{al}
a_l = \frac{(p-k)! }{p!}\,\left[ C_k^l \, A_k^l \, A_{p-k}^{k-l} \right]\;,
\ee
where $C_k^l$ are binomial coefficients $\binom{k}{l}$. We omit a combinatorial consideration that leads us to this formula. As a consistency check, one can show that  the coefficients satisfy the natural condition $\sum_{l=0}^k a_l  = 1$ meaning that we enumerated all possible permutations of originally symmetrized indices by re-organizing them in terms of the matrix elements \eqref{Inm}.\footnote{This can be directly seen by expressing $A^m_n = m! C^m_n$ and  using  the relation $ \sum_{i=0}^s C_{n}^i C_l^{s-i} = C_{n+l}^s$.}

Now, using explicit form of the matrix elements  \eqref{Inm} we find (up to an overall normalization)
\be\label{BlockintemrsCm}
F_{p,k} = \sum_{s=0}^{k}\;\;\; C_k^{s}\;\; (\mathbb{C}_0)^{k-s-1} 
\sum_{s \leq m_1 + ... +m_{s} \leq p-k}   
\mathbb{C}_{m_1}\mathbb{C}_{m_2} \cdots \mathbb{C}_{m_{s}}\mathbb{C}_{p-k-m_1-...-m_{s}}\;,
\ee
where factors $\mathbb{C}_n$ are given by \eqref{mathCn}. Expressing $\mathbb{C}_n$ in terms of the modular parameter $q$ the multiple summation can be reduced to just four sums. To this end, we split the multiple sum in two parts, which take into account  two terms in the last factor $\mathbb{C}_{p-k-m_1-...-m_{s}}$,
\be
F_{p,k} =q^{-\frac{p}{2}}\sum_{n=0}^{k}\;\;\; C_k^{n}\;\; (q-1)^{k-n-1}\bigg( q^{p-k+1} J_2(n,q)-J_1(n,q) \bigg)\;.
\ee
Here,
\be
\begin{aligned}
&J_1(n,q)=\sum_{n \leq m_1 + ... +m_{n} \leq p-k} (q^{m_1+1}-1)\cdots (q^{m_n+1}-1)=\sum_{r=0}^n (-)^{n+r} \,C^r_n \,q^r \tilde{J}_1(n,r,q) \,,\\
&J_2(n,q)=\!\!\!\!\sum_{n \leq m_1 + ... +m_{n} \leq p-k} \!\!\!\!q^{-\sum_{j=1}^n m_j}(q^{m_1+1}-1)\cdots (q^{m_n+1}-1)=\sum_{r=0}^n (-)^{n+r} \, C^r_n\,q^r \tilde{J}_2(n,r,q) \;,
\end{aligned}
\ee
and
\be
\tilde{J}_1(n,r,q)=\!\!\!\sum_{n \leq m_1 + ... +m_{n} \leq p-k} \!\!\!q^{-\sum_{j=1}^r m_j}\;,\qquad
\tilde{J}_2(n,r,q)=\!\!\!\sum_{n \leq m_1 + ... +m_{n} \leq p-k} \!\!\!q^{-\sum_{j=1}^n m_j-\sum_{j=1}^r m_j}\;.
\ee
We find
\be
\tilde{J}_1(n,r,q)=\sum _{i=n-r}^{(p-k) (n-r)} C_{i-1}^{n-r-1} \sum _{j=r}^{p-k-i} C_{j-1}^{r-1}\,q^j \;,
\ee
where the first binomial coefficient takes into account different ways to choose a set of $(n-r)$ elements $m_j$ not arising in the summand and the second coefficient
counts the restricted  compositions (see e.g. \cite{restr-comps}) for the set of the rest $r$ elements $m_j$.  We note that for $\tilde{J}_2(n,r,q)$ the evaluation differs only by 
interchanging the two sets and simultaneously replacing $q$ by $1/q$: 
\be
\tilde{J}_2(n,r,q)=\tilde{J}_1(n,n-r,1/q)\;.
\ee
This gives 
\be
\ba{l}
\dps
F_{p,k} = q^{-\frac{p}{2}} \sum _{n=0}^k  \,\sum _{r=0}^n\; (-1)^{r+k-1}\;  C_k^n\, C^r_n\, (1-q)^{k-n-1}\,\times
\\
\\
\dps
\hspace{10mm}\times\,\bigg( q^{p-k+1}\sum _{i=r}^{r (p-k)} C_{i-1}^{r-1} \sum _{j=n-r}^{p-k-i} C_{j-1}^{n-r-1}\,q^{r-j} -\sum _{i=n-r}^{(p-k) (n-r)} C_{i-1}^{n-r-1} \sum _{j=r}^{p-k-i}  C_{j-1}^{r-1}\, q^{r+j}\bigg)\;,
\ea\ee
that can be directly manipulated and after a somewhat tedious but straightforward re-summation yields the conformal block function \eqref{glob_poly}--\eqref{1ptcoefGLOBAL}.

Let us note the representation \eqref{BlockintemrsCm} has the triangle degree of complexity in the sense that it is simple when $k=0$ ($j_1=0$, the 0-point function, i.e. the character \eqref{fin_char}) or $k=p$ ($j_1 = 2j_p$, maximal admissible value of the external dimension \eqref{restrict})
\be
F_{_{p,0}}= \mathbb{C}_0^{-1} \mathbb{C}_p = q^{-\frac{p}{2}} \frac{1-q^{p+1}}{1-q}\;,
\qquad
F_{_{p,p}} = (\mathbb{C}_0)^{p}=  (-)^{p} q^{-\frac{p}{2}} (1-q)^{p}\;, 
\ee  
while the most complicated function  arises at $k=p/2$, when all multiple sums contribute.

\subsection{Character decomposition}
\label{sec:character}

The one-point block in the form \eqref{BlockintemrsCm} can be represented as a combination of zero-point blocks, i.e. characters, of various dimensions. To this end, we note that the character \eqref{fin_char} can be expressed  in terms of variables $\mathbb{C}_n$ as
\be
\chi_n  =  \frac{\mathbb{C}_n}{\mathbb{C}_0}\;, 
\ee
where, in its turn, $\mathbb{C}_0$ can be interpreted as the inverse character of the weight $\Delta=1/2$ representation, 
\be
\mathbb{C}_0 = -\frac{1}{\hat\chi_{\half}}\;, 
\qquad\text{where}\qquad
\hat\chi_{\half} = \frac{q^{\half}}{1-q}\;,
\ee
see \eqref{sl2_char}. Then, rewriting \eqref{BlockintemrsCm} in terms of the characters we arrive at the following  representation of the one-point block (up to a prefactor, recall that $p=2j_p$ and  $k=j_1$)
\be
\label{final}
F_{p,k} = \frac{1}{\left(\hat\chi_{\half}\right)^{k}}\,\sum_{s=0}^{k}\; \binom{k}{s}\;\;  
\sum_{s \leq m_1 + ... +m_{s} \leq p-k}   
\chi_{p-k-m_1-...-m_{s}} \prod_{i=1}^s \chi_{m_i}\;.
\ee

This form suggests that one can alternatively evaluate this expression  by using the Clebsch-Gordon rule for characters.  As an example, let the external dimension take the minimal admissible (bosonic) value, $j_1=k=1$. In this case, from  \eqref{final}  we obtain  the relation (recall that $p$ is even) 
\be
\label{Fp1_char}
F_{p,1} = \frac{1}{\hat\chi_{\half}}\,\sum_{m=0}^{\frac{p}{2}-1}\chi_m\chi_{p-m-1}\;.
\ee
Now we recall the Clebsch-Gordon series \eqref{fusion} in terms of the characters
\be
\chi_m \chi_{p-m-1} = \sum_{i=0}^m \chi_{p-2m-1+2i}\;.
\ee  
Substituting this expression into \eqref{Fp1_char} we find 
\be
\label{Fp1_char1}
F_{p,1} = \frac{1}{\hat\chi_{\half}}\, \sum_{n=0}^{\frac{p}{2}-1} \left(\frac{p}{2}-n\right)\chi_{p-2n-1} \;,
\ee
which after substituting  the explicit form of  characters in terms of $q$ gives back the one-point torus block function. 

To perform this calculation for general values of external dimension $j_1 = k$ one needs to know the Clebsch-Gordon series for tensoring any number $n$ of irreps of weights ${\bf j} = \{j_i = m_i/2\}$. It essentially reduces to knowing the Clebsch-Gordon numbers $N_j({\bf j})$ which are multiplicities of occurrence of a spin-$j$ in the range $[j_{min}, j_{max}]$, where min(max) weights are simply defined as $2j_{max} = \sum_{i=1}^n m_i$ and $2j_{min} = \sum_{i=1}^n (-)^{n+i-1}m_i$.\footnote{Alternatively, in order to evaluate $n$-fold tensor product one can apply the Clebsch-Gordon  procedure to each tensor couple of irreps to eliminate all tensor products in favour of direct sums.} However, a closed formula for the general Clebsch-Gordon numbers is an unsolved mathematical problem (for recent developments see e.g. \cite{Louck_2008}). 
This consideration leads to the following  representation\footnote{Note that the similar character decomposition is used to calculate the partition functions and correlators in $SU(N)$ lattice  gauge theories in the strong coupling regime, see e.g. review \cite{Caselle:2000tn}.} 
\be
\label{final_char}
\cF_{{j_p, j_1}}(q) = \frac{1}{\left(\hat\chi_{\half}\right)^{j_1}}\,\sum_{m\in D(j_p,j_1) }\; d_m\; \chi_{m}(q)\;,
\ee
which realizes the one-point block as the linear combination of characters. Here, unknown constant coefficients $d_m$ and the summation range $D(j_p,j_1)$ depend on the Clebsch-Gordon numbers $N_j({\bf j})$ for strings of characters  and factorial coefficients arising when re-summing multiple sums in the original formula \eqref{final}. An example is given by \eqref{Fp1_char1}.

\section{Two-point toroidal Wilson networks}
\label{sec:Two-point}

In what follows we represent two-point toroidal vertex functions of Section \bref{sec:toroidal} in terms of the symmetric tensor products along the lines of Section \bref{sec:one_point} and using the \3j Wigner symbols as in Section \bref{sec:wigner}.

\subsection{$s$-channel toroidal network}
\label{sec:2s}

Let us consider the toroidal vertex function \eqref{four_s} and insert resolutions of identities to express all ingredients as matrix elements in the standard basis (see Appendix \bref{sec:sl2})
\be
\label{four_s1}
\ba{l}
\dps
\stackrel{\circ}{V}_{\hspace{-1mm}{\rm (s)} \,b,e|a,c}(\tau, {\bf z}) 
=\sum_m\sum_n\sum_k\sum_l\sum_r
\,\Big(\langle j_b,m| \,W_b[0,2\pi\tau] \,|j_b, n\rangle\Big)\,\times
\\
\\
\dps
\hspace{12mm}\times \,\Big(\langle j_b, n|\,\I_{b; c, e}\,\,|j_e, k\rangle\otimes |j_c, l\rangle \Big)\,
\Big(\langle j_e,k| \I_{e; a, b}\,|j_b, m\rangle \otimes |j_a,r\rangle\Big) 
\,\langle j_c,l|\tilde c\rangle
\,\langle j_a, r|\tilde a\rangle\;,
\ea
\ee
where the last two matrix elements are given by coordinates of the tilded  boundary vectors,
\be
\label{four_s2}
\ba{c}
\dps
\langle j_c,l|\tilde c\rangle = \langle j_c,l| W_c[0,w_1] |\hat{\mathbb{lw}\,}\rangle_c \;,
\\
\\
\dps
\langle j_a,r|\tilde a\rangle = \langle j_a,r| W_a[0,w_2] |\hat{\mathbb{lw}\,}\rangle_a \;, 
\ea
\ee
where the transformed boundary  vectors are defined by expressions \eqref{trans1}, \eqref{trans2} in the respective representations. Now, we identify representations in two internal edges as $\cD_b \approx \cD_{j_{p_1}}$ and $\cD_e \approx \cD_{j_{p_2}}$, and two external edges as $\cD_c \approx \cD_{j_{1}}$ and $\cD_a \approx \cD_{j_{2}}$. The direct computation of the two-point torus blocks in $s$-channel from Wilson line networks according to \eqref{four_s1}--\eqref{four_s2} is given in Section~\bref{S2pt-proof}.

Following the discussion in Section  \bref{sec:one_point} we can also explicitly calculate each of the matrix elements entering \eqref{four_s1}--\eqref{four_s2}. The  Wigner $D$-matrix \eqref{M1} in the present case reads as
\be
\label{twoM1}
D_{\alpha_1\cdots\alpha_{\lambda_{p_1}}}^{\beta_1\cdots\beta_{\lambda_{p_1}}}=\frac{1}{\lambda_{p_1}!}\, D_{(\alpha_1}^{\beta_1}\cdots D_{\alpha_{\lambda_{p_1})}}^{\beta_{\lambda_{p_1}}} \;,
\ee 
where $\lambda_{p_1} = 2j_{p_1}$, while the projections of the boundary states are calculated as 
\be
\label{M1-2-3}
\tilde V^{(m)}_{\gamma_1\cdots\gamma_{\lambda_m}}=\tilde V^{(m)}_{\gamma_1}\cdots \tilde V^{(m)}_{\gamma_{\lambda_m}}\;, 
\qquad
m=1,2\;,
\ee      
where $\tilde V^{(1,2)}_\gamma$ are coordinates of the tilded transformed boundary vectors \eqref{four_s2}  in the basis of the fundamental representation, 
\be
\label{V12-tilde-a} 
\tilde V^{(m)}_\gamma=\delta_{\gamma,1}\exp{\frac{i w_m}{2}}+ \delta_{\gamma,2}\exp{-\frac{i w_m}{2}}\;,\qquad m=1,2\;,
\ee
cf. \eqref{V-tilde-a}. Now, we gather the above matrix elements together contracting them by means of the two intertwiners \eqref{proj}--\eqref{k}. Using the first intertwiner we obtain 
\be
\ba{l}
\dps
(W_1)^{\rho_1 ... \rho_{p_1}}_{\gamma_1 ... \gamma_{p_2}}=
\epsilon^{\alpha_{1}\beta_1}\cdots \epsilon^{\alpha_{\delta}\beta_{\delta}}\; 
D^{\rho_1 \ldots \rho_{\delta} \, \rho_{\delta+1} \ldots \rho_{\lambda_{p_1}}}_{\alpha_1 \ldots \alpha_{\delta} (\gamma_{1} \ldots \gamma_{\lambda_{p_1}-\delta}}
\;\tilde V^{(1)}_{\gamma_{\lambda_{p_1}-\delta+1}\ldots \gamma_{\lambda_{p_2}}) \beta_1 \ldots \beta_{\delta}}\;,
\ea
\ee
where $\delta=\frac{\lambda_{p_1}-\lambda_{p_2}+\lambda_1}{2}$. The second intertwiner gives
\be
\label{M123}
\ba{l}
\dps
(W_2)^{\rho_1 ... \rho_{p_1}}_{\gamma_1 ... \gamma_{p_1}}=
\epsilon^{\alpha_{1}\beta_1}\cdots \epsilon^{\alpha_{\varkappa}\beta_{\varkappa}}\; 
(W_1)^{\rho_1 \ldots \rho_{\varkappa}  \rho_{\varkappa+1} \ldots \rho_{\lambda_{p_1}}}_{\alpha_1 \ldots \alpha_{\varkappa} (\gamma_{1} \ldots \gamma_{\lambda_{p_2}-\varkappa}}
\;\tilde V^{(2)}_{\gamma_{\lambda_{p_1}-\varkappa+1}\ldots \gamma_{\lambda_{p_1}}) \beta_1 \ldots \beta_{\varkappa}}\;,
\ea
\ee
where $\varkappa=\frac{\lambda_{p_2}-\lambda_{p_1}+\lambda_2}{2}$. Finally, the overall contraction yields the two-point torus  block in the $s$-channel as
\be
\label{two-point-s-tensor}
\cF_s^{^{\Delta_{1,2}, \tilde \Delta_{1,2}}}\left(q, w_{1,2}\right) = (W_{2})_{\gamma_1\cdots\gamma_{\lambda_{p_1}}}^{\gamma_1\cdots\gamma_{\lambda_{p_1}}}\;.
\ee
The cylindrical coordinates on the torus can be related to the planar coordinates by $w  = i \log z $ so that the block function \eqref{glob-s} given in the planar coordinates is obtained from \eqref{two-point-s-tensor} by the standard conformal transformation for correlation functions. 
Explicit calculation of this matrix representation along the lines of the one-point block analysis in Section \bref{sec:one_point} will be considered elsewhere. Here, we just give one simple example with spins $j_{p_1} = j_{p_2}=j_{1}=j_{2}=1$ demonstrating that the resulting function is indeed  \eqref{glob-s}, see Appendix \bref{app:ex}.

\subsection{$t$-channel toroidal network} 
\label{sec:2t}

Let us consider the toroidal vertex function \eqref{four_t} and insert resolutions of identities to express all ingredients as matrix elements in the standard basis
\be
\label{four_t1}
\ba{l}
\dps
\stackrel{\circ}{V}_{\hspace{-1mm}{\rm (t)} \, c,e|a,b}(\tau, {\bf z})  
=\sum_m\sum_n\sum_k\sum_l\sum_r
\,\Big(\langle j_c,m| \,W_c[0,2\pi\tau] \,|j_c,n\rangle\Big)\,\times
\\
\\
\dps
\hspace{12mm}\times \,\Big(\langle j_c,n|\I_{c; c, e}\, |j_c,m\rangle\otimes|j_e,l\rangle\Big)\,
\Big(\langle j_e,l| \I_{e; a, b}\,|j_a,r\rangle\otimes|j_b,k\rangle\,\Big) 
\,\langle j_a,r|\tilde a\rangle
\,\langle j_b,k|\tilde b\rangle\;,
\ea
\ee
where the last two matrix elements are given by 
\be
\label{four_t2}
\ba{c}
\dps
\langle j_a,r|\tilde a\rangle = \langle j_a,r| W_a[0,w_1] |\hat{\mathbb{lw}\,}\rangle_a\;, 
\\
\\
\dps
\langle j_b,k|\tilde b\rangle = \langle j_b,k| W_b[0,w_2] |\hat{\mathbb{lw}\,}\rangle_b \;.
\ea
\ee
The representations are identified as $\cD_c \approx \cD_{j_{p_1}}$ and $\cD_e \approx \cD_{j_{p2}}$ for intermediate edges,  $\cD_a \approx \cD_{j_{1}}$ and $\cD_b \approx \cD_{j_{2}}$ for external edges. Using the matrix elements \eqref{twoM1} and \eqref{M1-2-3}, \eqref{V12-tilde-a} and then evaluating   the second intertwiner in \eqref{four_t1} we obtain 
\be
\ba{l}
\dps
(W_1)_{\gamma_1 ... \gamma_{p_2}}=
\epsilon^{\alpha_{1}\beta_1}\cdots \epsilon^{\alpha_{\delta}\beta_{\delta}}\; 
\tilde V^{(1)}_{\alpha_1 \ldots \alpha_{\delta} (\gamma_{1} \ldots \gamma_{\lambda_{1}-\delta}}
\;\tilde V^{(2)}_{\gamma_{\lambda_{1}-\delta+1}\ldots \gamma_{\lambda_{p_2}}) \beta_1 \ldots \beta_{\delta}}\;,
\ea
\ee
where $\delta=\frac{\lambda_{1}+\lambda_{2}-\lambda_{p_2}}{2}$. The first intertwiner gives
\be
\ba{l}
\dps
(W_2)^{\rho_1 ... \rho_{p_1}}_{\gamma_1 ... \gamma_{p_1}}=
\epsilon^{\alpha_{1}\beta_1}\cdots \epsilon^{\alpha_{\varkappa}\beta_{\varkappa}}\; 
D^{\rho_1 \ldots \rho_{\varkappa}  \rho_{\varkappa+1} \ldots \rho_{\lambda_{p_1}}}_{\alpha_1 \ldots \alpha_{\varkappa} (\gamma_{1} \ldots \gamma_{\lambda_{p_1}-\varkappa}}
\;(W_1)_{\gamma_{\lambda_{p_1}-\varkappa+1}\ldots \gamma_{\lambda_{p_1}}) \beta_1 \ldots \beta_{\varkappa}}\;,
\ea
\ee
where $\varkappa=\frac{\lambda_{p_2}}{2}$. Finally, the overall contraction yields the two-point torus  block in the $t$-channel \eqref{glob-t} in the cylindrical coordinates  as
\be
\label{two-point-t-tensor}
\cF_{t}^{_{\Delta_{1,2}, \tilde \Delta_{1,2}}}(q, w_{1,2}) = (W_2)_{\gamma_1\cdots\gamma_{\lambda_{p_1}}}^{\gamma_1\cdots\gamma_{\lambda_{p_1}}}\;.
\ee

Just as for $s$-channel blocks, we leave aside the straightforward check of this matrix representation and explicitly calculate just the simplest example given by spins $j_{p_1} = j_{p_2}=j_{1}=j_{2}=1$ to demonstrate that the resulting function is indeed  \eqref{glob-t}, see Appendix \bref{app:ex}.

\subsection{Wigner \3j symbol representation of the $s$-channel block}
\label{S2pt-proof}

Similarly to the one-point block the expansion coefficients of the two-point block in $s$-channel~\eqref{glob-s} with degenerate dimensions can be written in terms of hypergeometric functions
\be
\label{glob-s-1}
\begin{aligned}
&\cF_s^{^{\Delta_{1,2}, \tilde \Delta_{1,2}}}(q, z_{1,2}) =
q^{-j_{p_1}} z_1^{j_1-j_{p_1}+j_{p_2}} z_2^{j_2+j_{p_1}-j_{p_2}} \sum_{k=0}^{2j_{p_1}}\,\sum_{m=0}^{2j_{p_2}} f_{k,m}(j_{1,2}|j_{p_{1,2}})\, q^k\, \left(\frac{z_1}{z_2}\right)^{k-m},
\end{aligned}
\ee
where we used \eqref{degen} and\footnote{See eq. (2.13) in \cite{Alkalaev:2015fbw}.} 
\begin{align}
\label{fmn}
&f_{k,m}(j_{1,2}|j_{p_{1,2}})=
\frac{\tau_{k,m}(-j_{p_1},-j_1,-j_{p_2})\tau_{m,k}(-j_{p_2},-j_2, -j_{p_1})}{k!\, m!\,(-2j_{p_1})_k (-2j_{p_2})_m}\;,
\end{align}
where the $\tau$-coefficients are defined in~\eqref{A-tau}.
It can be written more explicitly in terms of generalized hypergeometric function as 
\begin{align}
&f_{k,m}(j_{1,2}|j_{p_{1,2}})=\frac{(j_{p_1}-j_{p_2}-j_1)_m (j_{p_2}-j_{p_1}-j_2)_k (j_{p_2}-j_{p_1}-j_1-m)_k (j_{p_1}-j_{p_2}-j_2-k)_m}{k! \,m!\,(-2j_{p_1})_k (-2 j_{p_2})_m  }\times\nonumber\\
&\hspace{10mm}\times\, _3F_2(2 j_{p_1}-k+1,\,-k,-m;\,j_{p_1}-j_{p_2}-j_2-k,\,j_{p_1}-j_{p_2}+j_2-k+1;\,1) \nonumber\\
&\hspace{10mm}\times \,_3F_2(2 j_{p_2}-m+1,-k,-m;j_{p_2}-j_{p_1}-j_1-m,j_{p_2}-j_{p_1}+j_1-m+1;1)\,\;.
\end{align}
On the other hand, the Wilson line network representation~\eqref{four_s1} reads 
\be
\label{four_s1_1}
\ba{l}
\dps
\stackrel{\circ}{V}_{\hspace{-1mm}{\rm (s)} \,b,e|a,c}(\tau, {w_{1,2}}) 
=\sum_m\sum_n\sum_k\sum_l\sum_r
\,\Big(\langle j_b,m| \,W_b[0,2\pi\tau] \,|j_b, n\rangle\Big)\,\times
\\
\\
\dps
\hspace{12mm}\times \,\Big(\langle j_b, n|\,\I_{b; c, e}\,\,|j_e, k\rangle\otimes |j_c, l\rangle \Big)\,
\Big(\langle j_e,k| \I_{e; a, b}\,|j_b, m\rangle \otimes |j_a,r\rangle\Big) 
\,\langle j_c,l|\tilde c\rangle
\,\langle j_a, r|\tilde a\rangle\;,
\ea
\ee
where $z_{1,2} = \exp{(-iw_{1,2})}$.
Or, using the notation introduced in Section~\ref{sec:wigner}, 
\be
\label{four_s1_2}
\ba{c}
\dps
\stackrel{\circ}{V}_{\hspace{-1mm}{\rm (s)} \,b,e|a,c}(\tau, {w_{1,2}}) 
=
\sum_m\sum_n\sum_k\sum_l\sum_r
 D^{(j_b)}{}^m{}_n [I_{b;c,e}]^n{}_{kl} [I_{e;a,b}]^k{}_{mr}\,V_{(j_c)}^l
\,V_{(j_a)}^r\;.
\ea
\ee
Substituting the Wigner D-matrix~\eqref{D-matrix}, the intertwiners in terms of \3j symbol \eqref{tri3-1}, and the two boundary vectors according to \eqref{four_s2},
\be
\label{6.2_new}
V_{(j)}^l = \langle j,l| W[0,w] |\hat{\mathbb{lw}\,}\rangle  = \beta_{j,l}\exp{(iwl)} \;,
\ee
we find 
\begin{align}
\label{four_s1_3}
&\stackrel{\circ}{V}_{\hspace{-1mm}{\rm (s)} \,b,e|a,c}(q, {w_{1,2}}) 
=\nonumber\\
&=\sum_{m,k,l,r} q^m \epsilon^{(b)}{}^{m,-k-l} \epsilon^{(e)}{}^{k,-m-r}
 \begin{pmatrix}
j_b & j_c &\, j_e \\
-k-l & l & \,k
\end{pmatrix} 
\begin{pmatrix}
j_e & j_a &\, j_b \\
-m-r & r & \,m
\end{pmatrix}
\beta_{j_c,l} e^{iw_2 l}\,\beta_{j_a,r} e^{iw_1 r}\;,
\end{align}
where the $\beta$-coefficients  are defined in~\eqref{trans2}, the sum over $n$ is removed by $\delta_n^m$ factor in the D-matrix~\eqref{D-matrix} and the \3j symbol $[W_{a,b,c}]_{mkn}$  property $m+n+k=0$ is used. The  Levi-Civita symbols $\epsilon^{(b)}{}^{m,-k-l}=(-)^{j_b-m}\delta_{m,k+l}$ and $\epsilon^{(e)}{}^{k,-m-r}=(-)^{j_e-k}\delta_{k,m+r}$ \eqref{levi_civita} remove other two summations 
\begin{align}
\label{four_s1_4}
&\stackrel{\circ}{V}_{\hspace{-1mm}{\rm (s)} \,b,e|a,c}(q, {w_{1,2}}) 
=\nonumber\\
&=\sum_{m, k} (-)^{j_b+j_e-m-k} q^m 
 \begin{pmatrix}
j_b & j_c &\, j_e \\
-m & m-k & \,k
\end{pmatrix} 
\begin{pmatrix}
j_e & j_a &\, j_b \\
-k & k-m & \,m
\end{pmatrix}
\beta_{j_c,m-k} e^{iw_2 (m-k)}\,\beta_{j_a,k-m} e^{iw_1 (k-m)}\;,
\end{align}
In order to compare the Wilson line network representation~\eqref{four_s1_1} with the CFT result~\eqref{glob-s-1} we need to identify representation labels as $a=j_2,\, b=j_{p_1},\, c =j_1,\, e = j_{p_2}$, and take into account the Jacobian 
 of the transformation to the planar coordinates $z_{1,2} = \exp{(-iw_{1,2})}$, see~\eqref{two-point-s-tensor-1},  
 \begin{align}
\label{four_s1_5}
&\hspace{-5mm}\cF_{s}^{_{\Delta_{1,2}, \tilde \Delta_{1,2}}}(q, z_{1,2}) 
=z_1^{j_1} z_2^{j_2} \sum _{k=-j_{p_1}}^{j_{p_1}} \sum _{m=-j_{p_2}}^{j_{p_2}} q^k (-)^{j_{p_1}+j_{p_2}-k-m} \left(\frac{z_1}{z_2}\right)^{k-m} \times\nonumber\\
&
\hspace{40mm}\times 
\beta_{j_1,k-m} \beta_{j_2,m-k}  
\begin{pmatrix}
\; j_{p_1} & j_1 & j_{p_2} \\
 -k & k-m & m \\
\end{pmatrix} 
\begin{pmatrix}
 j_{p_2} & j_2 & j_{p_1} \\
 -m & m-k & k \\
\end{pmatrix}.
\end{align}
Changing $k\rightarrow k+j_{p_1}$ and $m\rightarrow m+j_{p_2}$ we obtain 
\begin{align}
\label{four_s1_6}
\cF_{s}^{_{\Delta_{1,2}, \tilde \Delta_{1,2}}}(q, z_{1,2}) 
=
q^{-j_{p_1}} z_1^{j_1-j_{p_1}+j_{p_2}} z_2^{j_2+j_{p_1}-j_{p_2}} \sum _{k=0}^{2 j_{p_1}} \sum _{m=0}^{2 j_{p_2}} \tilde{f}_{k,m}(j_{1,2}|j_{p_{1,2}})\, q^k \left(\frac{z_1}{z_2}\right)^{k-m}\;,
\end{align}
where
\begin{align}
\label{ftmn}
&\hspace{-5mm}\tilde{f}_{k,m}(j_{1,2}|j_{p_{1,2}})=(-)^{-k-m}  \beta_{j_1,j_{p_2}-j_{p_1}+k-m}\,\beta_{j_2,j_{p_1}-j_{p_2}-k+m}\,\times\nonumber
\\
& 
\hspace{10mm}\begin{pmatrix}
 j_{p_1} & j_1 & j_{p_2} \\
 j_{p_1}-k\,\,\,\, &j_{p_2} -j_{p_1}+k-m \,\,\,\,& m-j_{p_2} \\
\end{pmatrix} 
\begin{pmatrix}
 j_{p_2} & j_2 & j_{p_1} \\
 j_{p_2}-m \,\,\,\,& j_{p_1}-j_{p_2}-k+m \,\,\,\,& k-j_{p_1} \\
\end{pmatrix}
.
\end{align}
Now, comparing~\eqref{four_s1_6} and~\eqref{glob-s-1}, we see that in order to verify the representation~\eqref{four_s1} we need to check 
 \be
 \label{fmn-ftmn}
 \tilde{f}_{k,m}(j_{1,2}|j_{p_{1,2}})=\varkappa_2 \, f_{k,m}(j_{1,2}|j_{p_{1,2}})\;,
 \ee
where the LHS is defined in~\eqref{ftmn} and the RHS in~\eqref{fmn} and $\varkappa_2$ is $(k,m)$-independent factor. Using the explicit representation~\eqref{3j-explicit} for \3j symbol in terms of the generalized hypergeometric function as well as  the following relation (see e.g.~\cite{prudnikov1986integrals}):
\begin{align}\label{Euler-transform-2}
&\, _3F_2(a,b,-k;c,d;1)=\nonumber\\
&(-)^k\, \frac{(d-a)_k (d-b)_k}{(c)_k(d)_k} \, _3F_2(1-d-k,a+b-c-d-k+1,-k;a-d-k+1,b-d-k+1;1)\,,
\end{align}
one can check that  the parameter $\varkappa_2$ is given by  
\begin{align}
&\varkappa_2 = 
\frac{2 \Gamma (2 j_{p_1}+1) \Gamma (2 j_{p_2}+1) }{\Gamma (j_1-j_{p_1}+j_{p_2}+1) \Gamma (j_2+j_{p_1}-j_{p_2}+1)}\sqrt{\frac{j_1\, j_2 }{\Gamma (j_1+j_{p_1}+j_{p_2}+2) \Gamma (j_2+j_{p_1}+j_{p_2}+2)}}\,\times\nonumber\\
&\times\sqrt{\frac{\Gamma (2 j_1) \Gamma (2 j_2) \Gamma (j_1-j_{p_1}+j_{p_2}+1) \Gamma (j_2+j_{p_1}-j_{p_2}+1)}{\Gamma (j_1+j_{p_1}-j_{p_2}+1) \Gamma (-j_1+j_{p_1}+j_{p_2}+1) \Gamma (j_2-j_{p_1}+j_{p_2}+1) \Gamma (-j_2+j_{p_1}+j_{p_2}+1) }}
\;.
\end{align}
Thus, the relation~\eqref{fmn-ftmn} holds which proves the  Wilson line network representation~\eqref{four_s1} for the two-point block in $s$-channel~\eqref{glob-s}.

A few comments are in order. First, we note that the general idea behind the proof is to observe that the $\tau$-function defining the block coefficients \eqref{fmn} can be expressed via the hypergeometric function $_3F_2$. On the other hand, one of the convenient representations of the Wigner \3j symbols is also given in terms of $_3F_2$. This ultimately allows comparing two dual representations of two-point configurations. Second, using this observation one can directly calculate the Wilson network representation for the $n$-point global torus block function  in the $s$-channel (also known as a necklace  channel) \cite{Alkalaev:2017bzx}.

\subsection{Wigner \3j symbol representation of the $t$-channel block}
\label{T2pt-proof}

In this section we follow the general calculation pattern elaborated in Sections \eqref{sec:wigner} and  \eqref{S2pt-proof}. To this end, we rewrite the two-point $t$-channel block~\eqref{glob-t} 
as\footnote{It can be shown that this factorized form of the block function is reduced to the product of two hypergeometric functions which are 1-point torus block and 4-point sphere $t$-channel block  \cite{Kraus:2017ezw}.}
\begin{align}
\label{global_t_p1}
\cF_{t}^{_{\Delta_{1,2}, \tilde \Delta_{1,2}}}(q, z_{1,2}) 
=
z_2^{-\Delta_1-\Delta_2} \sum _{m=0}^{\infty} g_{m}(\tilde \Delta_{1,2})q^{m+ \tilde \Delta_{1}}\sum _{l=0}^{\infty} h_{l}(\Delta_{1,2}| \tilde\Delta_{2})\, \left(\frac{z_1}{z_2}\right)^{l}\;,
\end{align}
where the coefficients
\be\label{g}
g_{m}(\tilde \Delta_{1,2})=\frac{\tau_{m,m}(\tilde\Delta_1, \tilde\Delta_2, \tilde\Delta_1)}{m!\,(2\tilde\Delta_1)_m} 
\ee
and
\be
\label{h}
h_{l}(\Delta_{1,2}| \tilde\Delta_{2})=(-)^{\tilde\Delta_2-\Delta_1-\Delta_2+l}\!\!\!\;\;\sum_{s=0}^\infty (-)^{s}
\binom{\tilde\Delta_2-\Delta_1-\Delta_2+s}{l}\,\frac{\sigma_{s}(\Delta_{1}, \Delta_{2}, \tilde\Delta_2)}{s!\,(2\tilde\Delta_2)_s}\;,
\ee
where to simplify the summation domain over parameter $s$ in the last formula we have adopted a formal rule that $\binom{x}{y}=0$ if $x<y$. All conformal dimensions are kept arbitrary and later on we choose those ones corresponding to the degenerate operators \eqref{degen}.

On the other hand, the Wilson line network representation~\eqref{four_t1} reads 
\be
\label{four_t1_new}
\ba{l}
\dps
\stackrel{\circ}{V}_{\hspace{-1mm}{\rm (t)} \, c,e|a,b}(\tau, {\bf z})  
=\sum_m\sum_n\sum_k\sum_l\sum_r
\,\Big(\langle j_c,m| \,W_c[0,2\pi\tau] \,|j_c,n\rangle\Big)\,\times
\\
\\
\dps
\hspace{12mm}\times \,\Big(\langle j_c,n|\I_{c; c, e}\, |j_c,m\rangle\otimes|j_e,l\rangle\Big)\,
\Big(\langle j_e,l| \I_{e; a, b}\,|j_a,r\rangle\otimes|j_b,k\rangle\,\Big) 
\,\langle j_a,r|\tilde a\rangle
\,\langle j_b,k|\tilde b\rangle\;.
\ea
\ee
Using the notation introduced in Section~\ref{sec:wigner}, 
\be
\label{four_s1_2}
\ba{c}
\dps
\stackrel{\circ}{V}_{\hspace{-1mm}{\rm (t)} \,b,e|a,c}(\tau, {w_{1,2}}) 
=
\sum_m\sum_n\sum_l\sum_k\sum_r
 D^{(j_c)}{}^m{}_n [I_{c;c,e}]^n{}_{ml} [I_{e;a,b}]^l{}_{rk}\,V_{(j_a)}^r
\,V_{(j_b)}^k\;.
\ea
\ee
Substituting the Wigner D-matrix~\eqref{D-matrix}, the intertwiners in terms of \3j symbol \eqref{tri3-1}, and the two boundary vectors \eqref{6.2_new}, we find 
\begin{align}
\label{four_t1_3}
&\stackrel{\circ}{V}_{\hspace{-1mm}{\rm (t)} \,b,e|a,c}(q, {w_{1,2}}) 
=\nonumber\\
&=\sum_{m,k,l,r} q^m \epsilon^{(c)}{}^{m,-m-l} \epsilon^{(e)}{}^{l,-r-k}
 \begin{pmatrix}
j_c & j_c &\, j_e \\
-m-l & m & \,l
\end{pmatrix} 
\begin{pmatrix}
j_e & j_a &\, j_b \\
-r-k & r & \,k
\end{pmatrix}
\beta_{j_a,r} e^{iw_1 r}\,\beta_{j_b,k} e^{iw_2 k}\;,
\end{align}
where the $\beta$-coefficients  are defined in~\eqref{trans2}. Now, we identify representation labels as $a=j_1$, $b=j_2$, $c =j_{p_1}$, $e=j_{p_2}$, and then use the Levi-Civita symbols and change to $z_{1,2} = \exp{(-iw_{1,2})}$,   
\begin{align}
\label{four_t1_4}
\cF_{t}^{_{\Delta_{1,2}, \tilde \Delta_{1,2}}}(q, z_{1,2})  
=z_1^{j_1} z_2^{j_2}\sum_{m=-j_{p_1}}^{j_{p_1}} (-)^{j_{p_1}+j_{p_2}-m} 
 \begin{pmatrix}
j_{p_1} & j_{p_1} &\, j_{p_2} \\
-m & m & \,0
\end{pmatrix}q^m \;\times
\nonumber\\
\times \sum_{k=-j_2}^{j_2} 
\beta_{j_1,-k} \,\beta_{j_2,k}\begin{pmatrix}
j_{p_2} & j_1 &\, j_2 \\
0 & -k & \,k
\end{pmatrix}
 \left(\frac{z_1}{z_2}\right)^k\;.
\end{align}
Changing $m\to m+j_{p_1}$ and $k\to l-j_1$ and assuming, for definiteness, that $j_1>j_2$ we obtain the  expression  
\begin{align}
\label{four_t1_5}
\cF_{t}^{_{\Delta_{1,2}, \tilde \Delta_{1,2}}}(q, z_{1,2})  
=z_2^{j_1+j_2}
\left[\sum_{m=0}^{2j_{p_1}} (-)^{j_{p_2}-m} 
 \begin{pmatrix}
j_{p_1} & j_{p_1} &\, j_{p_2} \\
j_{p_1}-m & m-j_{p_1} & \,0
\end{pmatrix}q^{m-j_{p_1}} \right]\;\times
\nonumber\\
\times \left[\sum_{l=j_1-j_2}^{j_1+j_2} 
\beta_{j_1,j_1-l} \,\beta_{j_2,l-j_1}\begin{pmatrix}
j_{p_2} & j_1 &\, j_2 \\
0 & j_1-l & \,l-j_1
\end{pmatrix}
 \left(\frac{z_1}{z_2}\right)^l\right]\;,
\end{align}
that has the following structure
\begin{align}
\label{four_t1_6}
\cF_{t}^{_{\Delta_{1,2}, \tilde \Delta_{1,2}}}(q, z_{1,2}) 
=
z_2^{j_1+j_2} \sum _{m=0}^{2 j_{p_1}} \tilde{g}_{m}(j_{p_{1,2}})q^{m-j_{p_1}}\sum _{l=j_1-j_2}^{j_1+j_2} \tilde{h}_{l}(j_{1,2}|j_{p_{2}})\, \left(\frac{z_1}{z_2}\right)^{l}\;.
\end{align}
Now, comparing the last relation ~\eqref{four_t1_6} and the original block function ~\eqref{global_t_p1} at integer negative weights \eqref{degen}, we see that in order to verify the representation~\eqref{four_t1} we need to check 
 \be
 \label{chi12}
 \tilde{g}_{m}(j_{p_{1,2}})=\chi_1 \, g_{m}(-j_{p_{1,2}})\quad 
 \text{and}\quad 
 \tilde{h}_{l}(j_{1,2}|j_{p_2})=\chi_2 \, h_{l}(-j_{1,2}|-j_{p_2}) \;,
 \ee
where the coefficients $\chi_{1,2}$ should not depend on $m,l$ and the coefficient functions $g_m,h_l$ are defined in~\eqref{g} and \eqref{h}. Note that $h_{l}(-j_{1,2}|-j_{p_2}) = 0$ if $l \notin [j_1-j_2,j_1+j_2]$.
Using the results of Section~\ref{sec:wigner} we get 
\begin{align}
\label{chi1}
\chi_1=
\frac{(-)^{j_{p_2}}(2j_{p_1})!}{\sqrt{(2j_{p_1}-j_{p_2})!\Gamma(2+2j_{p_1}+j_{p_2})}}
\;.
\end{align}
Similarly, using explicit representation for \3j symbol \cite{varshalovich} one can find that the second equality in~\eqref{chi12} holds with
 \begin{align}
\label{chi2}
\chi_2=\frac{(j_2-j_1-j_{p_2})_{j_{p_2}} \sqrt{\frac{\Gamma (2 j_2+1) \Gamma (j_1-j_2+1) \Gamma (j_1-j_2+j_{p_2}+1) \Gamma (j_{p_2}-j_1+j_2+1) \Gamma (j_1+j_2+j_{p_2}+2)}
{(-2 j_1)_{j_1+j_2} \Gamma (j_1+j_2-j_{p_2}+1)}}}{(-1)^{j_1+j_2+j_{p_2}} (-2 j_{p_2})_{j_{p_2}} \Gamma (2 j_{p_2}+1) \, _2F_1(-j_1-j_2-j_{p_2}-1,j_1-j_2-j_{p_2};-2 j_{p_2};1)}
\;.
\end{align}
Thus, we  conclude that the Wilson toroidal network operator ~\eqref{four_t1} does calculate the two-point torus block in $t$-channel~\eqref{glob-t}.


\section{Concluding remarks}
\label{sec:concl}
    
In this work we discussed toroidal Wilson networks in the thermal AdS$_3$ and how they compute $\sl2$ conformal blocks in torus CFT$_2$. We  extensively discussed the general formulation of the Wilson line networks which are actually $SU(1,1)$ spin networks, paying particular attention to key features that allow interpreting these networks as invariant parts of the conformal correlation functions, i.e. conformal blocks, on different topologies. We explicitly formulated toroidal Wilson line networks in the thermal AdS$_3$ and built corresponding vertex functions which calculate one-point and two-point torus conformal blocks with degenerate quasi-primary operators. In particular, both in the one-point and two-point cases we described two equivalent representations: the first is in terms of symmetric tensor products (multispinors), while  the second involves \3j Wigner  symbols. It turned out  that the calculation based on  the \3j symbols is obviously shorter than the one based on multispinors.  However, this is because the multispinor approach makes all combinatorial calculations manifest while using the \3j symbols we package this combinatorics into the known relations from the mathematical handbook.       

Our general direction for further research is to use the  spin network approach, which is a quite developed area (for review see e.g. \cite{Baez:1994hx}), in order to generalize Wilson line network representation of the $\sl2$ conformal blocks to the full Virasoro algebra $Vir$ conformal blocks. In this respect, recent papers \cite{Fitzpatrick:2016mtp,Besken:2017fsj,Hikida:2017ehf,Hikida:2018eih,Hikida:2018dxe} dealing with $1/c$ corrections to the sphere CFT$_2$ global blocks are interesting and promising. It would be tempting, for instance,   to formulate the  Wilson line network representation of quantum conformal blocks in $1/c$ perturbation theory for CFT$_2$ on general Riemann surfaces $\Sigma_g$. 

Obviously, this problem is quite non-trivial  already in the leading approximation since even global blocks on $\Sigma_g$ are unknown. In this respect one can mention the Casimir equation approach that characterizes global blocks as eigenfunctions of Casimir operators in OPE channels \cite{Dolan:2011dv}. As argued in \cite{Kraus:2017ezw} there are general group-theoretical arguments based on the gauge/conformal algebra relation \eqref{conf_trans} that force the Wilson network operators in the bulk to satisfy the Casimir equations on the boundary. It would be important to elaborate an exact procedure which identifies the Wilson  network operators with solutions of the Casimir equations for arbitrary OPE channels thereby showing the Wilson/block correspondence explicitly.           

Going beyond the global $c=\infty$ limit essentially leads to calculating multi-stress tensor correlators (as in the sphere CFT$_2$ case originally treated in \cite{Fitzpatrick:2016mtp} and \cite{Besken:2017fsj}).  However, the multi-stress tensor correlators on higher genus Riemann surfaces, and, in particular, on the torus, are quite complicated. In part, this is  due to non-trivial modular properties of double-periodic functions (in the torus case). 

In general, it might be that the Wilson line approach will prove efficient to calculate  block functions on arbitrary $\Sigma_g$  because the underlying spin networks are essentially the same as in the sphere topology case except for loops corresponding to non-trivial holonomies of the bulk space. It would be an alternative to the direct operator approach to calculating conformal blocks in CFT on Riemann surfaces.        

\paragraph{Acknowledgements.} The work  of K.A. was supported by the RFBR grant No 18-02-01024 and by the Foundation for the Advancement of Theoretical Physics and Mathematics “BASIS”.

\appendix

\section{$\sl2$ finite-dimensional representations}
\label{sec:sl2}

The commutation relations of $\sl2$ algebra are $[J_m, J_n] = (m-n) J_{m+n}$, where $m,n=0,\pm1$. Let $\cV_{\Delta}$ be a Verma module with weight (conformal dimension) $\Delta$. Its realization via the ladder operators reads as 
\be
\label{verma}
\ba{l}
J_0|\Delta,n\rangle = (\Delta+n)|\Delta,n\rangle\;,
\\
\\
J_{1}|\Delta,n\rangle = \sqrt{n(2\Delta+n-1)}|\Delta,n-1\rangle\;,
\\
\\
J_{-1}|\Delta,n\rangle = \sqrt{(n+1)(2\Delta+n)}|\Delta,n+1\rangle\;,

\ea
\ee
where $n=0,1,2,...$ enumerates the basis vectors $|\Delta,n\rangle$ and $|\Delta,0\rangle$ is the lowest-weight vector in $\cV_\Delta$. At negative weights  $\Delta = -j$, where $j\in \mathbb{Z}_+/2$ there is a singular vector at $n = -2\Delta+1 \equiv 2j+1$. The corresponding quotient module $\cD_j = \cV_{\Delta}/\cV_{-\Delta+1}$ is a non-unitary spin-$j$ representation of dimension $2j+1$.       It is spanned by vectors $|\Delta, n\rangle$ with $n=0,1,...,-2\Delta = 2j$. Keeping in mind the relation $\Delta= -j$ we call $j$ spin contrary to (conformal) weight $\Delta$.

The {\it standard} basis in $\cD_j$ is obtained from \eqref{verma} by redefining basis elements and introducing a new ``magnetic" parameter $m=n-j$ which runs $m = -j, -j+1,..., j-1,j$ so that the corresponding ladder operators now read
\be
\label{standard}
\ba{l}
J_0|j,m\rangle = m|j,m\rangle\;,
\\
\\
J_{1}|j,m\rangle = i\,\sqrt{(m+j)(m-j-1)}\,|j,m-1\rangle\equiv M_-(j,m)\,|j,m-1\rangle\;,
\\
\\
J_{-1}|j,m\rangle = -i\,\sqrt{(m-j)(m+j+1)}\,|j,m+1\rangle\equiv M_+(j,m)\,|j,m+1\rangle\;,

\ea
\ee
where we also defined coefficient functions $M_{\pm}(j,m)$. The inner product in $\cD_j$ is defined in the standard fashion as 
$\langle j,m|=|j,m\rangle^\dagger $,  
$J_0^\dagger = J_0$, $J_{\pm1}^\dagger = J_{\mp1}$.  The basis is orthonormal  $\langle j_1,m_1| j_2,m_2\rangle = \delta_{j_1,j_2}\delta_{m_1,m_2}$. 
From~\eqref{standard}  the explicit generators in the fundamental representation are
\be
J_0=\frac12
\left(
\begin{array}{cc}
 1 & 0 \\
 0 & -1 \\
\end{array}
\right)\;,
\quad
J_1=
\left(
\begin{array}{cc}
 0 & 0 \\
 -1 & 0 \\
\end{array}
\right)\;,
\quad
J_{-1}=
\left(
\begin{array}{cc}
 0 & 1 \\
 0 & 0 \\
\end{array}
\right)\;.
\ee
   
A representation  $\cD_j$ has both (highest)lowest-weight vectors. The corresponding highest-weight  and lowest-weight conditions are given by 
\be
\label{hw_lw}
\ba{l}
\text{HW}\;:\qquad J_{-1} |j,j\rangle=0\;, \qquad J_0 |j,j\rangle = j|j,j\rangle\;,
\\
\\ 
\text{LW}\;:\qquad J_{1} |j,-j\rangle=0\;, \;\;\;\;\;\;\; J_0 |j,-j\rangle = -j|j,-j\rangle\;.

\ea
\ee

\section{Tensor product of $\sl2$ spinors}
\label{sec:multi}

Using the Clebsch-Gordon series,  a representation $\cD_j$ of spin $j\in \mathbb{Z}/2$  can be realized as the symmetrized product $\cD_j =  (\cD_{\half}\otimes \cD_{\half})_{{\rm sym}}$, where $\cD_{\half}$ is the spinor (fundamental) representation. In components,   $T=T^{\alpha_1\cdots \alpha_\lambda} |e_{\alpha_1}\cdots e_{\alpha_\lambda}\rangle$,  where $T^{\alpha_1\cdots \alpha_\lambda}$ is a totally-symmetric multispinor with $\lambda = 2j$, spinor indices $\alpha = 1,2$, and $|e_{\alpha_1}\cdots e_{\alpha_\lambda}\rangle$ are basis elements. The highest-weight element is by construction $|h\rangle=|e_{1}\cdots e_{1}\rangle$.

A tensor product of representations $\cD_{j_1}$ and $\cD_{j_2}$ can be conveniently expressed in terms of multiplying two Young diagrams of respective lengths $\lambda_1 = 2j_1$ and $\lambda_2 = 2j_2$ as
\begin{figure}[H]
\hspace{2cm}\includegraphics[width=0.80\linewidth]{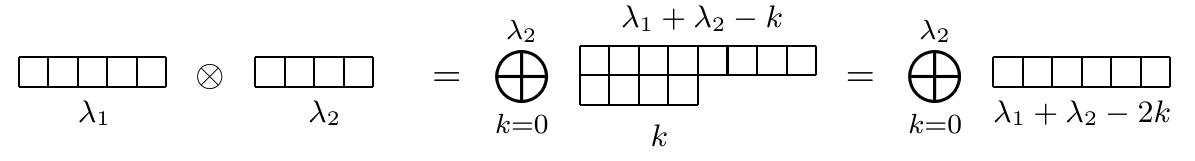}
\caption{The Clebsch-Gordon series \eqref{fusion} in terms of $\sl2$ spinor Young diagrams. }
\label{fig_univ_green}
\end{figure}

\noindent where, without loss of generality, we supposed that $\lambda_1\geq \lambda_2$. The last equality on Fig.  \bref{fig_univ_green} follows from the  fact that $2d$ Levi-Civita symbol $\epsilon^{\alpha\beta}=\;$\begin{picture}(12,12)(-1,7)
{\linethickness{0.210mm}
\put(00,15){\line(1,0){5}} 
\put(00,10){\line(1,0){5}} 
\put(00,05){\line(1,0){5}} 
\put(00,05){\line(0,1){10}} 
\put(05,05){\line(0,1){10}} 
}
\end{picture}
equals  a scalar by Hodge duality that lead to cutting off the two-row part of any $\sl2$ Young diagram.  

There are two technical tools that we use in the sequel. First, a totally symmetrized product of $\lambda$ spinors realizes   a rank-$\lambda$ multispinor,
\be
\label{spin_prod}
T^{\alpha_1 ... \alpha_{\lambda}} = \frac{1}{\lambda!}\, T_1^{(\alpha_1} \cdots T_\lambda^{\alpha_\lambda)}\;,
\ee   
or, in terms of ket vectors
\be
\label{ket_sym}
|e_{\alpha_1} \cdots e_{\alpha_{\lambda}}\rangle = \Big(|e_{\alpha_1}\rangle \otimes \cdots \otimes  |e_{\alpha_{\lambda}}\rangle\Big)_{{\rm sym}} \;.
\ee
Here, $|e_{\alpha}\rangle = \left\{|\half, \half\rangle, |\half, -\half\rangle \right\}$ are the standard basis elements of the spinor  $\sl2$ representation. The basis elements of $\cD_j$ parameterized by $\lambda = 2j$ read  
\be
|j,m\rangle \sim |e_{\alpha_1} \cdots e_{\alpha_{\lambda}}\rangle\;.
\ee

Second, the intertwiner $I_{j_3; j_1, j_2}$ or the projector on the rank $\lambda_3$ multispinor on the right-hand side of Fig. \bref{fig_univ_green} is defined   by contracting indices by the Levi-Civita symbols (cf. \eqref{tri3-1}) and subsequent symmetrization,   
\be
\label{proj}
R_{\alpha_1 ... \alpha_{\lambda_3}} = \epsilon^{\beta_1\gamma_1} \cdots \epsilon^{\beta_k\gamma_k} 
\,
S_{\beta_1... \beta_k(\alpha_1... \alpha_{\lambda_1-k}} 
\,
T_{\alpha_{\lambda_1-k+1}... \alpha_{\lambda_3}) \gamma_1... \gamma_k} \;,
\ee
where 
\be
\label{k}
k = \frac{\lambda_1+\lambda_2-\lambda_3}{2}\;.
\ee
The projector formula \eqref{proj} directly follows from the decomposition on Fig. \bref{fig_univ_green} by requiring that the right-hand side contains a representation with the weight $\lambda_3$. The two-row part of a given Young diagram gives a product of $k$ Levi-Civita symbols. The convention $\lambda_1 \geq \lambda_2$ guarantees the triangle inequalities  $\lambda_1-\lambda_2\leq \lambda_3 \leq \lambda_1+\lambda_2$ which define the summation domain in \eqref{fusion}, and so that $k\geq0$.

\section{One-point block via Legendre functions}
\label{appC}

The one-point torus block expressed in terms of the hypergeometric function \eqref{glob1pt} can be equally represented in terms of the Legendre functions (see e.g. \cite{AbrStegun}) as
\be
\label{legendre1}
\cF_{^{\tilde\Delta,\Delta}}(q) = (-)^{\tilde\Delta-\half}\,\Gamma(2\tilde\Delta)\;\frac{\;\;q^{\half}}{1-q}\; {\mathbb P}^{^{1-2\tilde\Delta}}_{_{-\Delta}} \left[\frac{1+q}{1-q}\right]\;. 
\ee 
Here, ${\mathbb P}^{^{\mu}}_{_\nu}[z]$ are the Legendre functions with  arbitrary parameters  $\mu, \nu \in \mathbb{C}$. For integer parameters $\mu, \nu \in \mathbb{Z}$ we get the associated Legendre polynomials which at $\mu=0$ become the standard Legendre polynomials.  

Using the hyperbolic parameterization with $x  = \half \log q= i \pi \tau$, where $\tau\in \mathbb{H}$ is the modulus, we can introduce yet another representation 
\be
\label{legendre2}
\cF_{^{\tilde\Delta,\Delta}}(q) = \half (-)^{\tilde\Delta+\half}\,\Gamma(2\tilde\Delta)\; \text{csh\,}  x  \;{\mathbb P}^{^{1-2\tilde\Delta}}_{_{-\Delta}} \left[-\coth{x}\right]\;. 
\ee 

A few comments are in order. (A) Due to the gamma function poles the above representations of the one-point torus block are not defined at $\tilde \Delta = -n/2$, where $n\in \mathbb{N}_0$. These particular dimensions define non-unitary finite-dimensional modules of (half-)integer weights. (B) At vanishing lower index ($\Delta=0$) the Legendre function reads
\be
{\mathbb P}^{^{\mu}}_{_0}[z] = \frac{1}{\Gamma(1-\mu)} \left[\frac{1+z}{1-z}\right]^{\frac{\mu}{2}}   
\ee
that can be used to show that in this case the one-point function \eqref{legendre1} goes into the $sl(2)$ character \eqref{sl2_char}. (C) The prefactor $\dps\frac{\;\;q^{\half}}{1-q}$ in \eqref{legendre1} is the character $\hat\chi_{\half}$ of the $sl(2)$ Verma module of the conformal weight $\half$.

\section{Explicit examples of two-point Wilson network operators }
\label{app:ex} 

{\bf Example I.} Choosing $\lambda_{p_1}=2$, $\lambda_{p_2}=2$, $\lambda_{1}=2$, $\lambda_{2}=2$ we explicitly evaluate  all quantities \eqref{twoM1}--\eqref{M123} to obtain 
\be
\label{twoM1e1}
D_{\alpha_1\alpha_2}^{\beta_1\beta_2}=D_{\alpha_1}^{\beta_1} D_{\alpha_{2}}^{\beta_{2}}+D_{\alpha_2}^{\beta_1} D_{\alpha_{1}}^{\beta_{2}} \;,
\ee 
\be
\label{M1-2-3e1}
\tilde V^{(1)}_{\gamma_1\gamma_{2}}=\tilde V^{(1)}_{\gamma_1}\tilde V^{(1)}_{\gamma_{2}}\;,
\qquad
\tilde V^{(2)}_{\gamma_1\gamma_{2}}=\tilde V^{(2)}_{\gamma_1}\tilde V^{(2)}_{\gamma_{2}}\;,
\ee  
\be
\ba{l}
\dps
(W_1)^{\rho_1\rho_{2}}_{\gamma_1\gamma_{2}}=
\epsilon^{\alpha_{1}\beta_1}\; 
D^{\rho_1\rho_{2}}_{\alpha_1 \gamma_{1}}
\;\tilde V^{(1)}_{\gamma_2}\tilde V^{(1)}_{\beta_1}+
\epsilon^{\alpha_{1}\beta_1}\; 
D^{\rho_1\rho_{2}}_{\alpha_1 \gamma_{2}}
\;\tilde V^{(1)}_{\gamma_1}\tilde V^{(1)}_{\beta_1}\;,
\ea
\ee
\be
\ba{l}
\dps
(W_2)^{\rho_1\rho_{2}}_{\gamma_1\gamma_{2}}=
\epsilon^{\alpha_{1}\beta_1}\; 
(W_1)^{\rho_1\rho_{2}}_{\alpha_1\gamma_{1}}
\;\tilde V^{(2)}_{\gamma_2}\tilde V^{(1)}_{\beta_1}
+
\epsilon^{\alpha_{1}\beta_1}\; 
(W_1)^{\rho_1\rho_{2}}_{\alpha_1\gamma_{2}}
\;\tilde V^{(2)}_{\gamma_1}\tilde V^{(1)}_{\beta_1}\;,
\ea
\ee 
where  $\tilde V^{(1,2)}_\gamma$ are given by \eqref{V12-tilde-a}. In the planar coordinates $z = \exp{(-iw)}$ the tilded boundary vectors are given by 
\be
\tilde V^{(m)}_\gamma=(z_m)^{-\half}\left(\delta_{\gamma,1}+ z_m\delta_{\gamma,2} \right)\;,\qquad m=1,2\;.
\ee
The two-point torus  block in the $s$-channel in the planar coordinates reads 
\be
\label{two-point-s-tensor-1}
\cF_s^{^{\Delta_{1,2}, \tilde \Delta_{1,2}}}(q, z_{1,2}) = (z_1)^{\Delta_1} (z_2)^{\Delta_2}(W_2)_{\gamma_1\gamma_{2}}^{\gamma_1\gamma_{2}}\;,
\ee
where the conformal weights $\Delta_1 =\Delta_2 = -1$, and the prefactors are the standard Jacobians relating correlations functions in different coordinates.   
Substituting all matrix elements into \eqref{two-point-s-tensor-1}, after tedious but elementary algebra, we find 
\be
\cF_s^{-1,-1,-1,-1} =\frac{q^2 z_1 (z_1-2 z_2)+q \left(z_1^2+z_2^2\right)+z_2 (z_2-2 z_1)}{2 q} \;,
\ee
which reproduces the $s$-channel block \eqref{glob-s} for the chosen conformal weights. 

{\bf Example II.} For the same weights  $\lambda_{p_1}=2$, $\lambda_{p_2}=2$, $\lambda_{1}=2$, $\lambda_{2}=2$ we obtain \eqref{twoM1e1}--\eqref{M1-2-3e1} as well as 
\be
\ba{l}
\dps
(W_1)_{\gamma_1\gamma_{2}}=
\epsilon^{\alpha_{1}\beta_1}\; 
\tilde V^{(1)}_{\alpha_1\gamma_{1}}
\;\tilde V^{(2)}_{\gamma_{2}\beta_1}+\epsilon^{\alpha_{1}\beta_1}\; 
\tilde V^{(1)}_{\alpha_1\gamma_{2}}
\;\tilde V^{(2)}_{\gamma_{1}\beta_1}\;,
\ea
\ee
\be
(W_2)^{\rho_1\rho_{2}}_{\gamma_1 \gamma_{2}}=
\epsilon^{\alpha_{1}\beta_1}\,
D^{\rho_1\rho_2}_{\alpha_1 \gamma_{1} }
\,(W_1)_{\gamma_{2} \beta_1}
+\epsilon^{\alpha_{1}\beta_1}\, 
D^{\rho_1\rho_2}_{\alpha_1 \gamma_{2} }
\,(W_1)_{\gamma_{1} \beta_1}\;.
\ee
The two-point torus  block in the $t$-channel \eqref{glob-t} in the planar coordinates is
\be
\label{two-point-t-tensor}
\cF_{t}^{_{\Delta_{1,2}, \tilde \Delta_{1,2}}}(q, z_{1,2}) =  (z_1)^{\Delta_1} (z_2)^{\Delta_2}(W_2)_{\gamma_1\gamma_{2}}^{\gamma_1\gamma_{2}}\;.
\ee
Substituting all matrix elements into \eqref{two-point-t-tensor} we find that 
\be
\cF_{t}^{-1,-1,-1,-1} =
\frac{(q-1)^2 \left(z_1^2 + 4 z_1 z_2 + z_2^2\right)}{6 q} \;,
\ee
which reproduces the $t$-channel block \eqref{glob-t} for the chosen conformal weights.



\providecommand{\href}[2]{#2}\begingroup\raggedright\endgroup

\end{document}